\documentclass[aps,pre,twocolumn,groupaddress]{revtex4}
\usepackage{amsmath}
\usepackage{amsfonts}
\usepackage{amssymb}
\usepackage{epsfig}
\usepackage{graphicx}

\newcommand{\be}{\begin{equation}}
\newcommand{\ee}{\end{equation}}
\renewcommand{\phi}{\varphi}
\newcommand{\ba}{\begin{eqnarray}}
\newcommand{\ea}{\end{eqnarray}}

\begin{document} 

\title{Intermittent dynamics and logarithmic domain growth 
during the spinodal decomposition of a glass-forming liquid}

\author{Vincent Testard}
\affiliation{Laboratoire Charles Coulomb, 
UMR 5221 CNRS and Universit\'e Montpellier 2, 34095 Montpellier, France}

\author{Ludovic Berthier}
\affiliation{Laboratoire Charles Coulomb, 
UMR 5221 CNRS and Universit\'e Montpellier 2, 34095 Montpellier, France}

\author{Walter Kob}
\affiliation{Laboratoire Charles Coulomb, 
UMR 5221 CNRS and Universit\'e Montpellier 2, 34095 Montpellier, France}

\date{\today}
\begin{abstract}
We use large-scale molecular dynamics simulations of a simple
glass-forming system to investigate how its liquid-gas phase separation
kinetics depends on temperature. A shallow quench leads to a fully
demixed liquid-gas system whereas a deep quench makes the dense
phase undergo a glass transition and become an amorphous solid.
This glass has a gel-like bicontinuous structure that evolves very
slowly with time and becomes fully arrested in the limit where thermal
fluctuations become negligible. We show that the phase separation kinetics
changes qualitatively with temperature, the microscopic dynamics evolving
from a surface tension-driven diffusive motion at high temperature to a
strongly intermittent, heterogeneous and thermally activated dynamics
at low temperature, with a logarithmically slow growth of the
typical domain size. These results shed light on recent experimental
observations of various porous materials produced by arrested spinodal
decomposition, such as nonequilibrium colloidal gels and bicontinuous
polymeric structures, and they elucidate the microscopic mechanisms
underlying a specific class of viscoelastic phase separation.

\end{abstract}



\maketitle

\section{Introduction}
The behavior found in nonequilibrium kinetic phenomena such as
self-assembly~\cite{kapral}, pattern formation~\cite{hohenberg2}, and
phase ordering kinetics~\cite{bray} is typically much richer than the
one encountered in equilibrium since a broader range of morphologies and
a more complex relaxation dynamics can be observed in systems that are
far from equilibrium. While this has been an active research area since
many decades~\cite{hohenberg2,hohenberg}, the field has seen a surge
of interest in the last few years, since progress in the synthesis of
colloidal particles with complex shapes and tunable pairwise interactions
permits the self-assembly of materials of ever growing complexity,
i.e. systems that exist only if one masters their intricate formation
process~\cite{kapral}.

One of the most important phenomenon occurring in nonequilibrium systems
is spinodal decomposition~\cite{bray,furukawa2}. The case of shallow
temperature quenches has been studied in great detail using theory as
well as experiments. Various regimes for the time evolution of the
average domain size have been predicted, and observed in controlled
experiments and simulations of simple fluids using a multitude of
computational approaches~\cite{koch,lebo,laradji,cates,das1}. However, a
much greater complexity can be expected if the dense phase is not a simple
liquid~\cite{hansen} but is itself a ``complex'' fluid~\cite{witten}. Of
particular importance is the so-called ``viscoelastic phase
separation''~\cite{tanaka,protein} which is characterized by a strong
physical (mainly, rheological) asymmetry between the two coexisting
phases. For instance, one can study the phase separation between two
fluids of unequal viscosities, or the coexistence between a solid and
a fluid phase. In the following we shall be concerned with a situation
intermediate between these two, where one component will evolve from
being a simple fluid to become a highly viscous liquid or an amorphous
glass phase, thus giving rise to the phenomenon of a {\it glass-gas
phase separation}~\cite{letter,jackle,danchinov,varrato_12}.

The phase ordering process between a fluid and an amorphous solid
can be expected to display a complex phenomenology, since even the
equilibrium bulk behavior of (homogeneous) amorphous glasses is not
well understood~\cite{binder_2011,rmp}. When suddenly quenched from high
temperature to below the glass temperature, a glass-forming material
evolves slowly with time, undergoing a nonstationary aging dynamics
characterized by intermittent, heterogeneous dynamics occurring far
from equilibrium~\cite{rmp,bouchaud_96,kob_97,buisson_03,luca2}. It is
not clear how this aging glass state will evolve if it is given by the
dense phase in the complex bicontinuous structure formed during spinodal
decomposition. Since a glass behaves mechanically like a solid, one can
expect that the bicontinuous structure formed after the quench into the
coexistence region becomes rigid, and as a result will become kinetically
arrested into a bicontinuous porous structure~\cite{jackle}. In that
case, the glass-gas phase separation would thus be a conceptually very
simple way of producing porous media. However, one can also expect
that the aging of the glass phase enables intermittent, thermally
activated microscopic rearrangements, which could potentially make the
porous material very fragile. To advance our understanding regarding
these questions we present here the results of our study concerning the
interplay between phase separation kinetics and aging behavior resulting
from deep quenches at low temperatures in a simple glass-forming model.

Although our main motivation is to obtain a fundamental understanding
of the glass-gas phase separation kinetics, there are also several
experimental considerations that motivate such a study. Firstly, a
number of colloidal systems and protein solutions can be modelled as
spherical particles with nearly hard-core repulsion and longer-ranged
attraction, whose range and strength can be controlled. Thus, they
will undergo a phase separation in some part of the phase diagram,
which might possibly interfere with the colloidal glass formation
occurring at higher densities. Therefore, the idea that, at least in
some materials, gelation results from a kinetically arrested glass-gas
spinodal decomposition has been explored in several experimental
works~\cite{weitz,lu,lu2,cardinaux,gibaud2,gibaud,bristol,paul,paddy}.
This process has also been the subject of a number of numerical studies,
mostly aimed at reproducing realistic coarse-grained pair interactions
characterizing colloidal systems that were specifically studied
experimentally~\cite{lu2,foffi,dave,dave2,zacca,araki,sciortino_11}.
More recent studies, in line with our own work~\cite{letter},
have considered more diverse systems, such as the phase separation
kinetics of a coarse-grained model for buckyball C$_{60}$ carbon
molecules~\cite{paddyc60}.  Moreover, bicontinuous disordered structures
reminiscent of the ones obtained in spinodal decompositions may also
be found in colorful bird feathers, and were recently interpreted as
incompletely phase separated polymeric glasses~\cite{dufresne,dufresne2},
thus demonstrating the broad relevance of the problem of the
glass-gas phase separation. Finally, even more complex mixtures
that are relevant to food processing~\cite{food} or solar cell
technology~\cite{solar1,solar2,solar3} also exhibit kinetically arrested
spinodal decompositions, that might result from the fact that one
component becomes mechanically rigid.

There are several possible ways to study coarsening processes
by means of theory~\cite{bray}. One efficient approach is to study a
coarse-grained model of a biphasic material using a Ginzburg-Landau
free energy functional of the two-phase system complemented with
phenomenological dynamical equations to incorporate relevant dynamical
and mechanical properties of each phase~\cite{tanaka,onuki}.  For simple
fluids, this amounts to studying model $H$ in the classification
scheme proposed by Hohenberg and Halperin~\cite{hohenberg}.  Possible
extensions to viscoelastic materials have indeed been considered in
the past~\cite{tanaka}, and numerically integrated to obtain insights
into some specific viscoelastic phase separation processes. Still, it
remains difficult to faithfully incorporate in this approach the complex
(typically highly nonlinear and history dependent) physical properties
of real glass-forming materials in their aging regime.  Moreover,
such coarse-grained equations cannot provide direct information on the
microscopic dynamics responsible for the evolution of such bicontinuous
materials.

To avoid this drawback we use here a microscopically realistic
description of the homogeneous glass using an atomistic model combined
with molecular dynamics techniques. We study its behavior during phase
separation over a broad range of control parameters, mainly density and
temperature.  Indeed, numerous successful atomic-scale simulations of
the simpler situation of a liquid-gas spinodal decomposition have been
reported~\cite{koch,yamamoto,velasco,laradji}.  In Ref.~\cite{evans}, a
Lennard-Jones system was quenched to low temperature in the coexistence
region, and the resulting crystal-gas phase separation was studied,
but no kinetic arrest was reported (see Ref.~\cite{crystalPRL} for a
recent related experimental study).  Simulations of realistic colloidal
interactions have been reported in Refs.~\cite{foffi,dave,zacca,araki}, but the
quenches have been performed to very low temperatures and hence particles
aggregate nearly irreversibly and thermal fluctuations play little
role, which corresponds effectively to zero-temperature quenches in our
approach. Thus, a careful study of the crossover regime between ordinary
spinodal decomposition and irreversible aggregation is so far not available.

In this work, we fill this gap and provide a detailed numerical study
of the phase separation kinetics between a gas and a glass-forming
material at various temperatures encompassing the glass transition of the
bulk material. We describe how the phase separation kinetics changes
qualitatively with temperature, the microscopic dynamics evolving from
the well-known diffusive motion driven by surface tension for shallow
quenches, to a qualitatively different coarsening regime in which dynamics
becomes strongly intermittent, spatially heterogeneous and thermally
activated at low temperature, leading to logarithmically slow growth
of the typical domain size. A short account of our results has been
published~\cite{letter}.

Our paper is organized as follows.  In Sec.~\ref{model} we define the
model, provide technical details about our numerical simulations, and
discuss the phase diagram of the system and the relevant parameters
to be explored in this work.  In Sec.~\ref{qualitative} we provide
a qualitative description of the temperature influence on the
spinodal decomposition kinetics.  In Sec.~\ref{structure} we present
several structural characterizations of the bicontinuous structures and
in Sec.~\ref{temporal} we discuss the time evolution of these structures,
characterizing in detail the influence of temperature on the growth law.
In Sec.~\ref{dynamics} we provide insights into the microscopic mechanisms
responsible for the coarsening structures at high and low temperatures.
In Sec.~\ref{competition} we discuss the nature of the coexistence line
below the glass transition temperature, that was the subject of a recent
controversy in the literature.  In Sec.~\ref{conclusion}, we close the
paper with some perspectives for future work.

\section{Model and phase diagram}

In this section, we describe details of the Lennard-Jones model used
in this study and provide some technical informations about the numerical
simulations. We then describe the relevant features of the phase diagram
of the model. Finally we introduce a coarse-grained density field that
will be useful to the analysis of the simulations.

\label{model}

\subsection{Model and technical details}

To study the interplay between liquid-gas phase separation and the
liquid-glass transition we consider a simple Lennard-Jones model for a
liquid that was first devised to study the dynamics of glass-forming
materials in the bulk~\cite{KA}. The model is a 80:20 binary mixture
of Lennard-Jones particles with asymmetric interaction parameters
chosen such that the minority component frustrates, and
therefore efficiently prevents, the crystallization of the majority
component. The details of the interaction parameters are as in the
original publication~\cite{KA}.  In the following we use Lennard-Jones
units corresponding to the majority component, expressing length in units
of the particle diameter, $\sigma$, and time in units of $\tau = \sqrt{m
\sigma^2 / \epsilon}$, where $m$ is the particle mass and $\epsilon$
the energy scale of the Lennard-Jones interaction between particles of
the majority component.

We integrate Newton's equation of motion using LAMMPS~\cite{plimpton}
for $N$ particles enclosed in a volume $V$, working with periodic
boundary conditions. We work at constant number density, $\rho = N/V$,
and adjust the temperature $T$ using periodic velocity rescaling as a
simple thermostatting procedure. The equations of motion are integrated
using a standard velocity Verlet scheme, using a discretization time
step of $0.01$ in reduced Lennard-Jones units.

To study the kinetics of phase separation, we first prepare homogeneous
samples at the desired density $\rho$, working at high temperature,
$T=3.0$, well above the critical point $T_c \approx 1.2$, until thermal
equilibrium is reached.  We then instantaneously quench the temperature
$T$ to the desired final value, where the phase separation dynamics
starts. In the following we will denote as the ``age'' of the system the
time elapsed since the quench to the final temperature. To improve the statistics
of our quantitative measurements and the robustness of our findings,
we have repeated simulations at each state point using 5 to 10 samples,
using independently prepared samples at high temperatures.

Specific attention has been paid to system size. While numerical
studies of the glass transition in the homogeneous liquid typically
require simulating about $N=10^3$ particles, we have found that up to $N=10^6$
particles are needed to obtain results devoid of finite size effects
during the phase separation. We have obtained most of our quantitative results
using $N = 3 \cdot 10^5$ particles. Where appropriate, we will discuss
the $N$-dependence of our numerical results.

\subsection{Coarse-graining the density}

\label{subcoarse}

Since the bicontinuous structures produced during a spinodal
decomposition are characterized by a typical length scale that is often
much larger than the typical particle size, the local structure of the
fluid is largely irrelevant. Therefore, it will prove useful to first
coarse-grain the density field before quantifying the spatial fluctuations
of the obtained field~\cite{das2,das3}. This coarse-grained density will
also be used to facilitate the visualizations of the particle configurations.

We start from the microscopic density field, which is defined as 

\be
\label{micro}
\rho_{\rm micro} 
({\bf r}) = \sum_{i=1}^N \delta ({\bf r} - {\bf r}_i ),
\ee

\noindent
where ${\bf r}_i$ is the position of particle $i$.
Our first step is to discretize space by dividing it into boxes of
linear size $\xi_{\rm b}$, so that continuous space is now replaced by
a discrete lattice containing $V/\xi_{\rm b}^3$ sites.  We consider
a discrete density, $\rho ({\bf r})$, defined for discrete positions
${\bf r}$ located at the center of the boxes described above, as

\be
\label{rho}
\rho ( {\bf r} ) = \frac{3}{4 \pi \xi_{\rm s}^3} \sum_{i=1}^N 
\theta ( \xi_{\rm s} - |{\bf r} - {\bf r}_i|),  
\ee

\noindent
where $\theta(x)$ is the Heaviside function and a second coarse-graining
length, $\xi_{\rm s}$, is introduced.  Thus, the density at position
${\bf r}$ takes into account all particles located in a sphere of radius
$\xi_{\rm s}$ centered at ${\bf r}$.

Finally, we obtain the desired coarse-grained density $\bar \rho ({\bf
r})$ by using the following weighted average of $\rho({\bf r})$ over
the boxes surrounding ${\bf r}$:

\begin{eqnarray}
\label{barrho}
\bar{\rho}({\bf r}) & = & \frac{1}{8} \big[ 2 \rho({\bf r}) 
+ \rho({\bf r} + \xi_b {\bf e}_x) +  \rho({\bf r} + \xi_b {\bf e}_y) 
+ \rho({\bf r} +  \xi_b {\bf e}_z) \nonumber \\
& & + \rho({\bf r} - \xi_b {\bf e}_x) +  \rho({\bf r} - \xi_b {\bf e}_y) 
+ \rho({\bf r} -  \xi_b {\bf e}_z)
 \big]. 
\end{eqnarray}

\noindent
Here ${\bf e}_\alpha$ is a unit vector in direction $\alpha$.

The procedure described by Eqs.~(\ref{rho}, \ref{barrho}) is easy to
implement. It returns for each lattice site a coarse-grained density
field which is a much smoother function than the microscopic density field
$\rho_{\rm micro}({\bf r})$ in Eq.~(\ref{micro}). The two coarse-graining
length scales $\xi_{\rm b}$ and $\xi_{\rm s}$ can be adjusted by seeking a
compromise between having a smooth field without losing too much spatial
resolution. After having tried several combinations~\cite{these}, we
have settled to the values $\xi_{\rm b} = \sigma / 2$ and $\xi_{\rm s}
= \sigma$, so that the lattice spacing is equal to the particle radius,
while the density field is coarse-grained by taking into account the
immediate neighbourood of each particle.

\begin{figure}
\includegraphics[width=85mm]{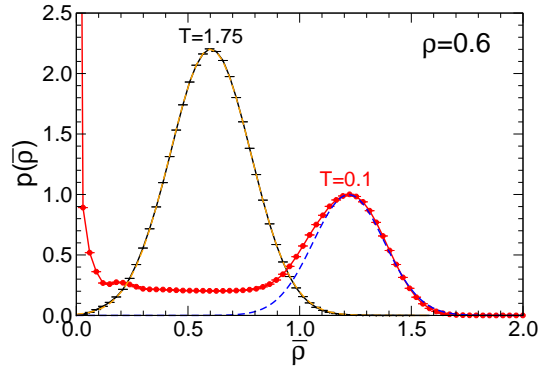}
\vspace*{-12mm}
\caption{\label{density} 
Probability distribution function of the coarse-grained density field
for $\rho=0.6$ for both a homogeneous fluid configuration at $T=1.75$
and for a phase separating system at $T=0.1$. The dashed lines are a fit 
with a Gaussian.}
\end{figure}

Having defined the coarse-grained density field $\bar{\rho}({\bf
r})$, we can now easily transform a set of particle coordinates from a
single particle configuration into the probability distribution of the
coarse-grained density, $p(\bar \rho)$.  In Fig.~\ref{density}, we show
this distribution for two configurations obtained at density $\rho=0.6$.
The first example is measured for $T=1.75$, where the system is in the
homogeneous fluid phase. As shown by the dashed line, the distribution
is well described by a Gaussian functional form with a maximum located
at $\bar \rho = 0.6$, as expected. More interesting is the second case
at $T=0.1$ where the system is in the phase coexistence region. For
this low temperature, complete phase separation is not reached. The
distribution $p(\bar \rho)$ directly reflects this phase coexistence,
since it is characterized by two peaks. One peak is located at very low
density, representative of the gas phase (whose very small density is
not resolved in the scale chosen in Fig.~\ref{density}).  A second peak
corresponds to the dense phase and has a maximum near $\rho \approx 1.2$
in this specific example. In between these peaks, the distribution is
nearly flat. We checked that this intermediate density band corresponds to
sites located near the interfaces between the two phases where the density
can take any value comprised between the ones of the gas and fluid phases.

These observations can be used for two purposes: First, by adjusting
the peak at high density to a Gaussian distribution, we can directly
measure the average density of the dense phase and follow its evolution
during a quench to the coexistence region. This will be used for instance
in Sec.~\ref{competition} to determine the coexistence line at low
temperature where complete phase separation is not reached at long times,
see also the binodal curve in the phase diagram of Fig.~\ref{phase}.

A second application is the possibility to clearly distinguish,
using $\bar \rho$ and $p(\bar \rho)$, between cells that belong to either
of the two phases, and those belonging to the interfaces. To this end,
we need to determine a threshold density delimiting the dense phase from
the gas phase.  By careful visual inspections of several configurations
at various state points, we have chosen $\bar \rho =0.42$ as giving the
most faithful representation of the particle configurations.  Thus, we
define from now on cells with $\bar \rho > 0.42$ as belonging to the fluid
phase, and cells with $\bar \rho < 0.42$ as those forming the gas phase.
Interfaces are represented by cells that are in the dense phase and 
that have at 
least one neighboring cell that is {\it not} in the dense phase.
We have used for instance these definitions to produce the images shown
in Figs.~\ref{image} and \ref{movie}, where only cells belonging to the
interfaces were shown, using different colors for gas and fluid phases.

\subsection{Phase diagram}

\label{subphase}

\begin{figure}
\includegraphics[width=85mm]{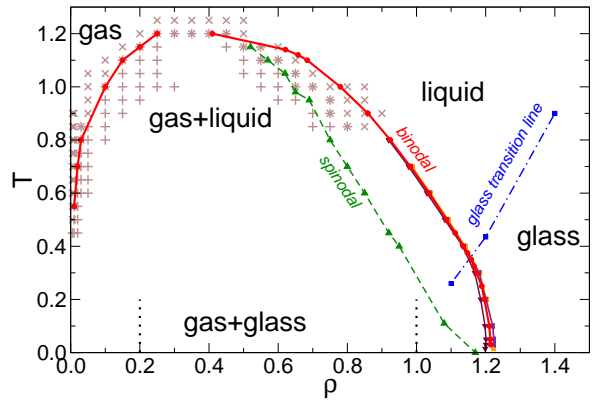}
\vspace*{-7mm}
\caption{\label{phase} 
Temperature-density phase diagram of the binary Lennard-Jones mixture,
showing the fluid, liquid and glass homogeneous phases.  The ``glass
transition line'' has been obtained from a mode-coupling analysis of the
glassy dynamics in the liquid phase~\cite{gillesmct}.  The ``binodal''
line separates homogeneous from biphasic states. It is obtained
by direct inspection of the state points marked by symbols at high
temperature, and from the evaluation of the density in the dense phase
of biphasic states at low temperature.  The ``spinodal'' line is taken
from  Ref.~\cite{sastry}. }
\end{figure}

Using a combination of direct visual inspection of equilibrium
configurations coupled to more quantitative methods to analyze the
morphology of biphasic atomistic configurations (as described above),
we have determined the phase diagram shown in Fig.~\ref{phase}, which
we complement with some relevant literature data. In this phase diagram,
the control parameters are the temperature, $T$, and the number density,
$\rho$.

When density is high enough, roughly $\rho \geq 1.2$, the system
is always homogeneous. Similarly, the system is a homogeneous fluid
at high temperature, $T \geq T_c \approx 1.2$, which corresponds
to the critical temperature. At large density and temperature, the
system is a simple liquid, but its dynamics slows down dramatically
as temperature decreases, without showing sign of crystallization.
Therefore, the system undergoes a glass transition from a viscous liquid
to an amorphous glass as temperature is decreased at constant density.
This process has been extensively studied before~\cite{KA}. To set
the typical temperature scale for the glass formation, we include in
Fig.~\ref{phase} the density dependence of the temperature obtained by
analyzing the dynamics in the framework of the mode-coupling theory of
the glass transition~\cite{gillesmct}.  This temperature was determined
by a power-law fit to the growth of the equilibrium relaxation time.
It is known to represent a useful temperature scale below which
it becomes difficult to reach thermal equilibrium in a standard
numerical simulation. Thus, for all practical purposes the system is
in a homogeneous glass phase below the ``glass transition line'' shown in
the phase diagram of Fig.~\ref{phase}.

For temperatures below $T_c \approx 1.2$ and low enough densities, $\rho
\leq 1.2$, the homogeneous system is unstable and phase coexistence is
observed, see Fig.~\ref{phase}.  To determine the shown binodal line 
we have performed quenches to a number of state points
(symbols in the figure) and analyze whether the system remains homogeneous
at very long times, i.e. we determined whether or not the distribution of
coarse-grained density discussed in the previous subsection has only one
peak. Very close to the binodal, we payed attention to metastability and
hysteretic effects and performed additional numerical tests to assess the
location of the binodal~\cite{these}. Unfortunately, this direct method
becomes inefficient if temperature becomes small, $T \leq 0.4$, because
complete phase separation does not occur in the limited time window of
our simulations, and hysteresis effects become more pronounced at lower
temperatures. In this region, therefore, the coexistence line has been
determined by performing quenches well within the coexistence region,
and by measuring the average density of the dense phase of the biphasic
configurations, using the method described above in Sec.~\ref{subcoarse}.
We have made sure that both methods yield consistent results in the
vicinity of  $T=0.4$ where we switch from one approach to the other.

Let us mention here that there exists a controversy in the literature
about the nature and the behavior of the coexistence line below its
intersection with the bulk glass transition line. While one group
claims that the binodal line is little affected by its intersection
with the glass transition line~\cite{lu,lu2}, another group reports
that the density dependence of the binodal evolves non-monotonically
with decreasing temperature, and becomes slaved to the glass transition
line at low temperatures~\cite{cardinaux,gibaud2,gibaud}.  We shall
specifically come back to this point below, and so we describe 
here only briefly our own findings in Fig.~\ref{phase}, which are somewhat
intermediate between the two situations described in the literature. While
the binodal line we have determined is not slaved to the glass transition
line, its temperature dependence is nevertheless clearly affected by
the intersection with this line, so that the measured
binodal becomes nearly independent of density at low temperatures,
as shown in Fig.~\ref{phase}.

\section{Qualitative overview of results}

In this section we present a general overview of the distinct types of
morphologies obtained in the course of our numerical studies. We then
describe qualitative aspects of the kinetics of the phase separation
process at various state points.

\label{qualitative}

\subsection{Biphasic morphologies at long times}

\begin{figure}
\includegraphics[width=85mm]{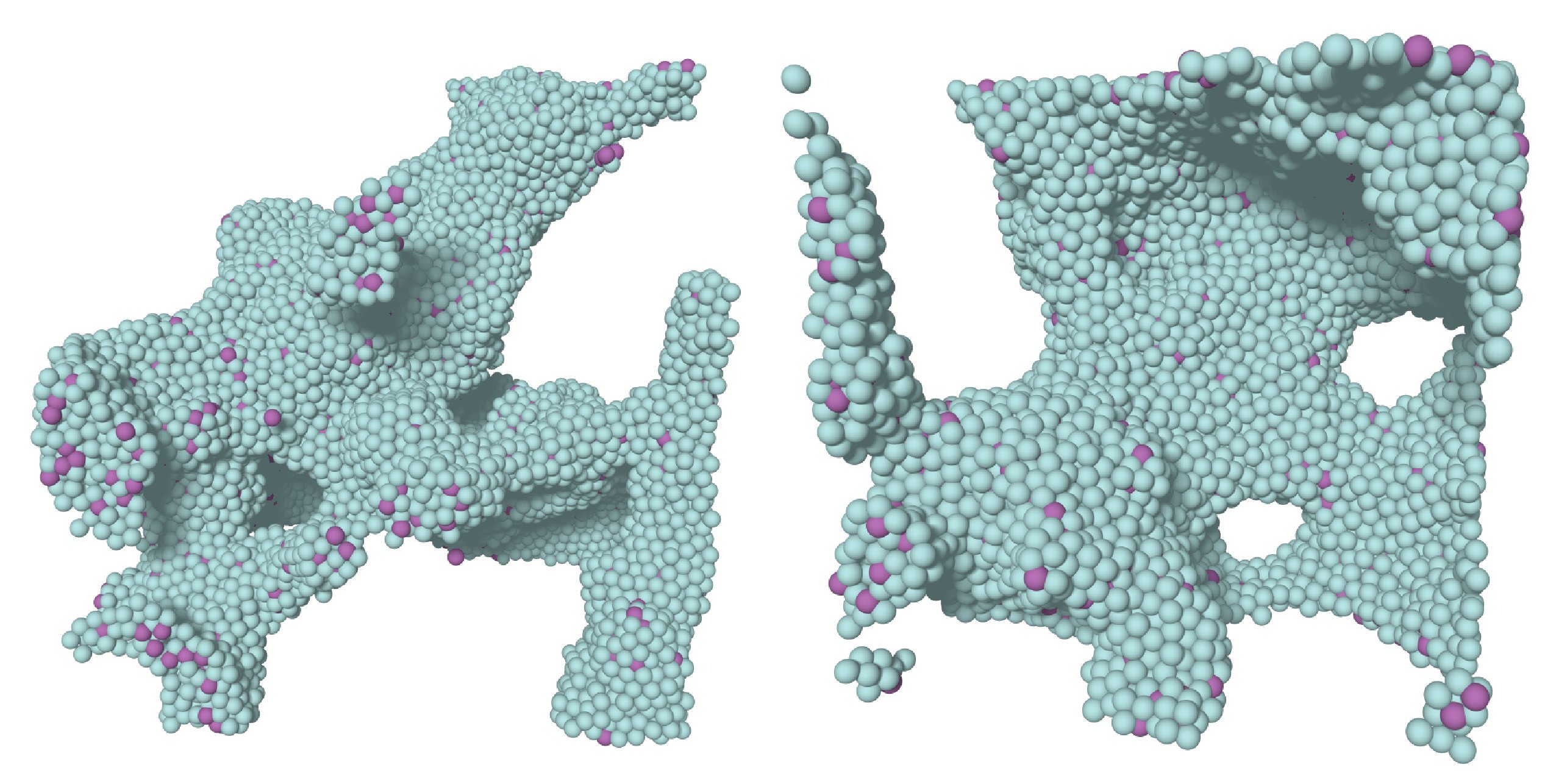}
\caption{\label{gels} 
Snapshots of representative bicontinuous gel-like configurations obtained
at long time, $t=10^4$, for $T=0.1$ and densities $\rho=0.2$ (left)
and $\rho=0.6$ (right).  For the sake of clarity, these snapshots show
only a small fraction (about 16~\%) of the total number of particles.}
\end{figure}

\begin{figure}
\includegraphics[width=85mm]{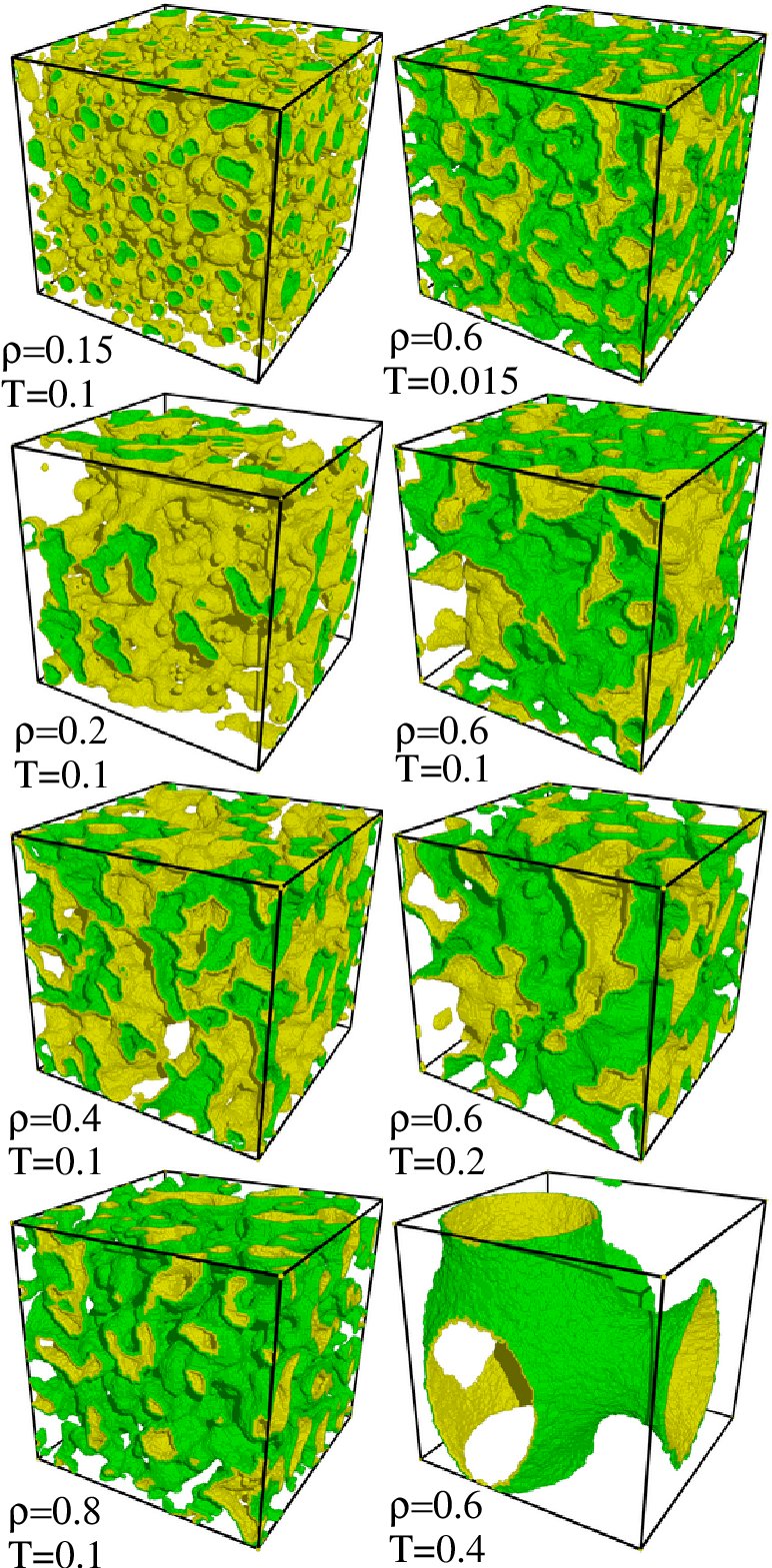}
\caption{\label{image} 
Representative configurations obtained at long times, $t=10^4$,
and various state points indicated in the figure. Left: constant
low temperature, $T=0.1$, and various densities.  Right: constant
intermediate density, $\rho=0.6$, and various temperatures. For all
snapshots, periodic boundary conditions are used, as is obvious for
instance in the bottom right configuration.}
\end{figure}

We start by discussing the morphologies observed when we quench the
system to various state points in the coexistence region of the phase
diagram in Fig.~\ref{phase}.  Figures~\ref{gels} and \ref{image} show
that for a broad range of densities, $0.2 \leq \rho \leq 1.0$, and for
low enough temperatures, $T \leq 0.1$, the morphologies obtained at
long times after the quench to the final temperature, $t=10^4$, are
bicontinuous gel-like structures. These particle configurations are
strongly reminiscent of the nonequilibrium colloidal gels observed for
instance using confocal microscopy~\cite{lu,bristol}.  By changing the
density and the temperature, the typical length scales characterizing
these porous structures change and a central goal of the present paper
is to quantify these changes.

As can be noticed from the snapshots shown in Fig.~\ref{gels}, it is not
trivial to visualize the complex morphologies of porous, bicontinuous
materials in three dimensions. Therefore, to ease the visualization,
we have implemented a numerical method to localize the two phases,
and notably the interfaces between them. The procedure relies on the
definition of a coarse-grained density field, as explained above in
Sec.~\ref{subcoarse}. While we primarily developed this procedure to
quantitatively characterize the numerically obtained bicontinuous
structures (described below), we find that it is also useful for
visualization purposes, as demonstrated in Fig.~\ref{image} where now
the entire simulation box is shown and the geometries are more 
easily visualized than by showing particle configurations directly.

In the left column of the figure we show representative configurations
obtained at long times for a low temperature, $T=0.1$ (recall that
the critical temperature is $T_c \approx 1.2$) and various densities.
For low density, $\rho=0.15$, one obtains disconnected droplets of dense
fluids that slowly coarsen with time. For $\rho \geq 0.2$, a bicontinuous
structure is obtained, with a dense phase which occupies an increasing
fraction of the total volume as density increases from $\rho=0.2$, to
$\rho=0.4$ and $\rho=0.8$.  For $\rho \geq 1.0$ (not shown), it is the
low-density gas phase which now occupies disconnected bubbles inside
the dense phase.

In the right column of Fig.~\ref{image} we show the evolution of
the final morphology obtained in our simulation for $t=10^4$ for
quenches at fixed intermediate density, $\rho=0.6$, and different
temperatures. While the phase separation proceeds rapidly for shallow
quenches, $T \geq 0.5$, it is not complete for a deeper quench at $T=0.4$,
see Fig.~\ref{image}. (Note that the fact that in this panel the final
configuration is not the expected spherical object might also be
a finite size effect.)  Decreasing further the temperature, $T \leq
0.2$, one observes that even at the end of our simulations an intricate
bicontinuous structure remains apparent, indicating that phase separation
is far from being reached. Even smaller domains are obtained at large
times for very low temperatures, $T=0.015$.  It is therefore clear that
at low $T$ the phase separation kinetics is slowed down dramatically
and thus the typical domain size remains relatively modest even at very
long times. In the subsequent sections of the paper we will characterize
these observations in a quantitative manner.

\subsection{Kinetics of phase separation}

\begin{figure*}
\includegraphics[width=40mm]{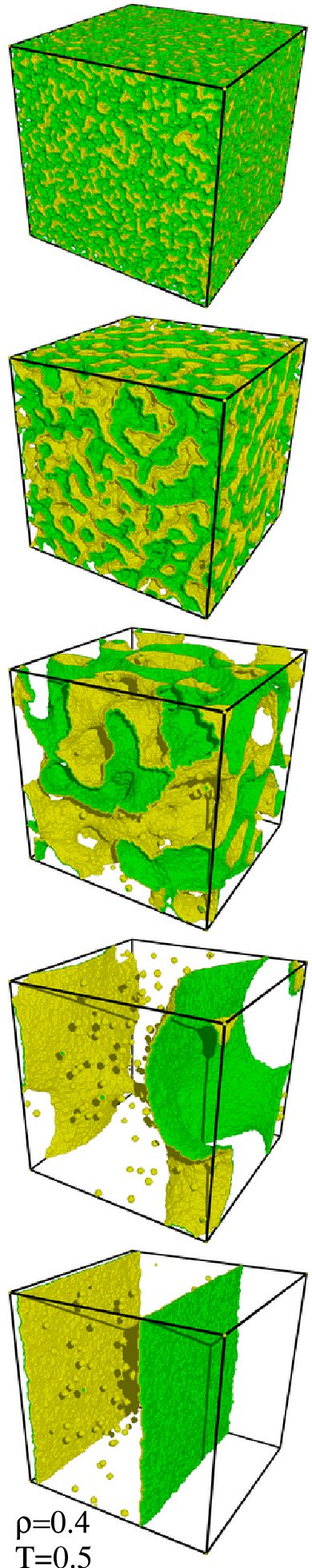}
\includegraphics[width=40mm]{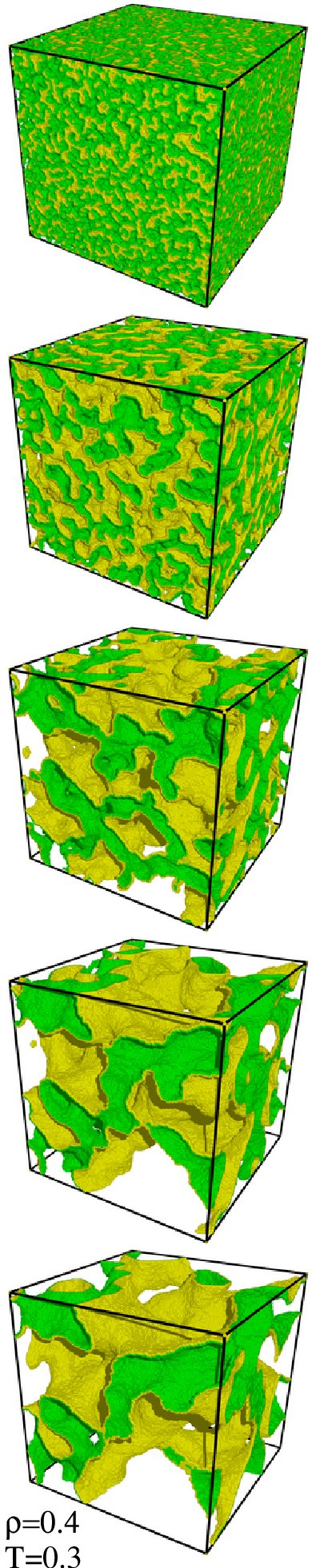}
\includegraphics[width=40mm]{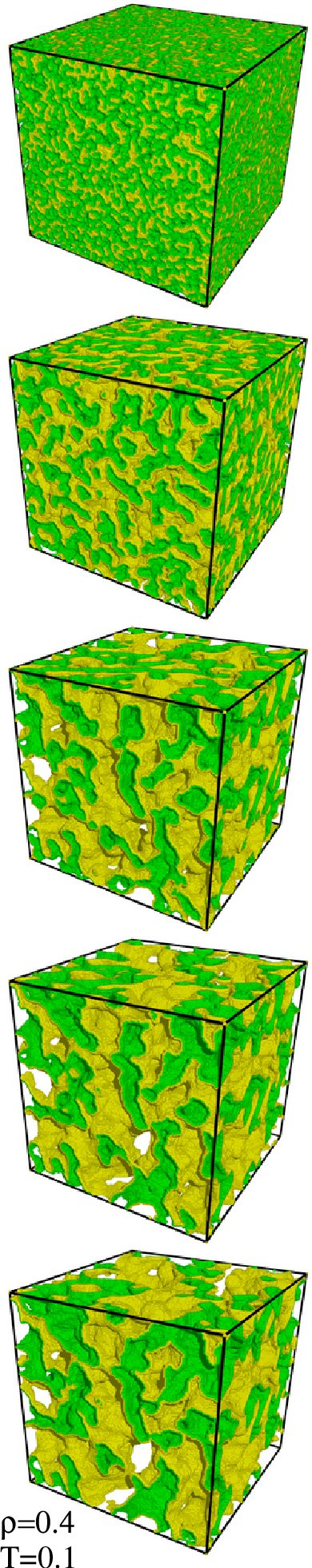}
\includegraphics[width=40mm]{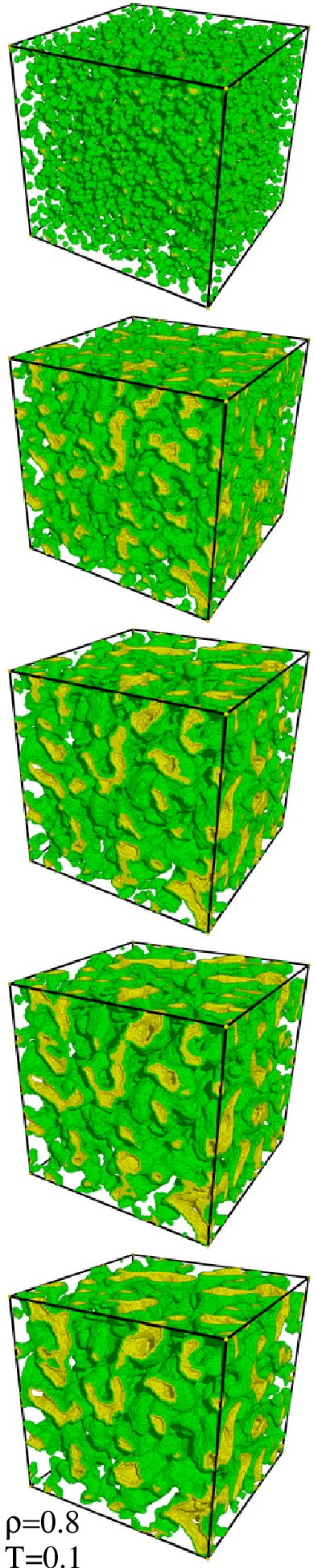}
\caption{\label{movie}
Time evolution of the phase separating system after a temperature quench 
at various state points indicated in each column. The selected timescales
are identical in the four column. From top to bottom, $t=1.45$, 
13.2, 120, 1096, and $10^4$.}
\end{figure*}

We now make a qualitative description of the kinetics of the phase
separation process at various state points, as illustrated in the
time series shown in Fig.~\ref{movie}. Each snapshot in a given row is
separated from the previous one by a factor of 10 in time.

The first column in Fig.~\ref{movie} reproduces the typical time
evolution observed in molecular dynamics studies of the liquid-gas
spinodal decomposition~\cite{koch,das1,yamamoto}.  For this density,
$\rho=0.4$, the gas and liquid phases occupy similar volumes. The
temperature is $T=0.5$, about half the critical temperature $T_c$,
and thus the quench is not very deep. Using the phase diagram in
Fig.~\ref{phase}, we see that the liquid phase is still well above
the glass transition line. It thus behaves as a simple liquid, and so
the spinodal decomposition takes place with no interference from the
physics related to the glass transition. Indeed, one observes that a
bicontinuous structure forms over a very short time, $t \approx 10$,
and then coarsens slowly with time. For a finite system as the one
used in our numerical simulations, the phase separation proceeds until
a simple geometry is reached with a flat interface separating the two
phases. For an infinite system, of course, spinodal decomposition would
proceed indefinitely in a scale-invariant manner~\cite{bray}.

When temperature is decreased to $T=0.3$ as in the second column in
Fig.~\ref{movie}, the initial stages of the phase separation process are
little affected and similar bicontinuous structures are formed at early
times. This is expected since dynamics at short times mainly results
from the well-known spinodal instability~\cite{kapral}: The homogeneous
state being fully unstable, a density modulation develops with a dominant
wavevector. Therefore, temperature plays little role in this initial
process.  However, the coarsening dynamics which follows is now clearly
affected by the temperature, since the final configuration is no longer
a fully demixed system but remains a complex bicontinuous structure. That
the coarsening slows down is reasonable since the self-diffusion constant
of the dense liquid phase decreases when temperature is lowered, and is
expected to be already very small at $T=0.3$~\cite{KA}. On the basis of
these snapshots alone, it cannot be decided whether the phase separation
kinetics changes nature, or is simply slowed down by a trivial factor in
time which could for instance be absorbed in a rescaling $t$ with the
diffusion constant. We shall see later that such a simple rescaling is
insufficient.

At even lower temperature, $T=0.1$ (third column in Fig.~\ref{movie}),
the situation is much less ambiguous: While the phase separation for
early times, $t \leq 10^2$, proceeds as before, the slowing down is
now dramatic, as demonstrated by the fact that the two snapshots for
$t=10^3$ and $t=10^4$ are virtually identical (at least to the eye).
This implies that at this temperature phase separation is nearly arrested
at intermediate times, and well before the typical domain size has reached
the system size. Finally the fourth column in Fig.~\ref{movie} illustrates
that a similar slowing down of the phase separation is observed for a
broad range of densities, the example shown being $\rho=0.8$ at $T=0.1$
where again the last two panels look identical but are separated by one
order of magnitude in timescales. These observations show that at low
temperatures, the domain growth is strongly slowed down, and indicate
that the physics of the coarsening process at low temperatures cannot be 
explained by a simple rescaling of the time scale.

Below, we will establish that the phase separation is in fact not fully
arrested at low temperature, but that instead it has changed nature
in the sense that the observed growth law depends qualitatively on
temperature. Another point that will be carefully considered is whether
the near-arrest observed in the low temperature images of Fig.~\ref{movie}
results from a finite size effect, or if it survives in the thermodynamic
limit.

\section{Structural analysis}
\label{structure}

To quantify the above qualitative observations we must first
characterizate the observed bicontinuous structures in terms of
quantitative observables and then determine how these depend on time and
temperature.  In this section we introduce and compare several structural
indicators, and show that the so-called ``chord length distribution''
represents an efficient choice to describe phase separating systems in
our particle-based numerical simulations.

\subsection{Pair correlation function}

Since we know at each timestep of the simulation the position of all
the particles, an obvious choice of a microscopic function to quantity
the large scale structures is to record
the pair correlation function, defined as~\cite{hansen}

\be
g({\bf r},t) = \frac{1}{\rho N} \sum_{j=1}^N \sum_{k=1}^N 
\left\langle \delta( {\bf r} - ({\bf r}_j(t) - {\bf r}_k(t) )) \right\rangle,  
\label{gofr}
\ee

\noindent
where the brackets represent an average over independent initial
conditions and trajectories, and $ {\bf r}_j(t)$ is the position of
particle $j$ at time $t$.  Since our configurations are isotropic
we further perform a spherical average and divide by the phase space
factor $4 \pi r^2$ to obtain $g(r = |{\bf r}|,t)$.  We have considered
also partial pair correlation functions, specializing the sums in
Eq.~(\ref{gofr}) to either one of the two species of the binary Lennard-Jones
mixture.  However, since we are interested in the large-scale structure
of the configurations, we shall only discuss the total pair correlation
described by Eq.~(\ref{gofr}). Note that the pair correlation function
is frequently measured in colloidal experiments using confocal microscopy
techniques and hence is a quantity that is experimentally accessible.

\begin{figure}
\includegraphics[width=100mm]{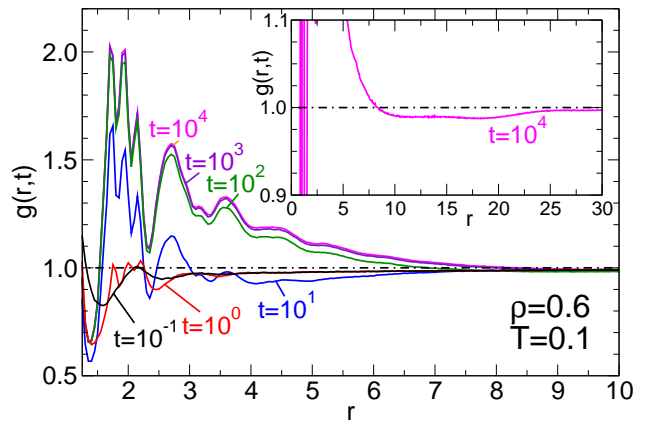}
\vspace*{-15mm}
\caption{\label{gr} Time evolution of the pair correlation function
$g(r,t)$ for a quench at $\rho=0.6$ and $T=0.1$,
and various logarithmically separated times. The inset shows a zoom of
the small amplitude oscillation around $g(r,t)\approx 1$ which can serve as
a measure of the average domain size (see main text for details).}
\end{figure} 
  
In Fig.~\ref{gr} we show examples of the time evolution of $g(r,t)$
for a quench to $\rho=0.6$ and  $T=0.1$.  By definition, $g(r,t)$ is
proportional to the probability to find a particle at distance $r$ from
a particle located at the origin.  Therefore $g(r,t)$ describes for short
distances the local amorphous structure of the dense phase.  Accordingly,
it shows a pronounced first peak corresponding to inter-particle distances
(occurring at $r \approx 1.0$ and not shown in the figure), followed by
quickly decaying and smooth oscillations representative of an amorphous
structure that has no long-range crystalline order.

The difference with a homogeneous liquid appears at larger distances:
While in a homogeneous liquid $g(r,t)$ rapidly converges to unity
if $r$ increases, for the heterogeneous phase separating systems it
shows oscillations around unity even at large distances, as shown in
Fig.~\ref{gr}.  If $t$ is large, the first oscillation below 1 represents
the average distance between a particle taken at random in a dense domain
to a neighboring gas region. This physical interpretation suggests that
a possible quantitative definition of the average domain size, $L(t)$,
can be obtained from $g(L(t),t)=1$. The data shown in Fig.~\ref{gr},
however, indicate that the oscillations of $g(r,t)$ around unity at large
$r$ have a rather small amplitude, in particular if $t$
is large (see Inset in Fig.~\ref{gr}). Our simulations have shown that
the above definition is physically sensible, in that it coincides well
with the typical domain size seen in the snapshots~\cite{these}. However,
we also noticed that this measurement is prone to very large statistical
fluctuations, since the amplitude of the oscillations in $g(r,t)$ is very
small. As a result, the use of this method to follow the time evolution
of the phase separating systems does require a very large numerical
effort since highly accurate pair correlation functions must be measured.

Additional visual inspections indicate that the large distance behavior
of $g(r,t)$ is in fact strongly influenced by a small number of very
large domains found in the system, whose statistical properties strongly
fluctuate from one run to another. Therefore, although this two-point
function is a simple structural correlation which corresponds also to the
quantity usually analyzed in theoretical calculations, our work suggests
that, at least for particle-based numerical simulations, it does not
represent the most practical choice to determine the average domain size.

\subsection{Structure factor}

\begin{figure}
\includegraphics[width=85mm]{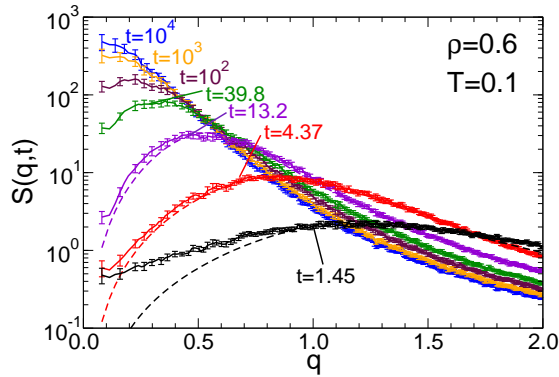}
\vspace*{-11mm}
\caption{\label{sq} 
Time evolution of the static structure factor, 
Eq.~(\ref{sqt}), for a quench at $\rho=0.6$ and $T=0.1$, and various
logarithmically separated times. The lines are fits using 
Eq.~(\ref{furu}).}  
\end{figure}

Since the static structure factor is the Fourier transform of the pair
correlation function~\cite{hansen}, it carries {\it a priori} the same
physical content. It is, however, more easily accessible to experiments
using for instance light scattering techniques.  It is defined as

\be
S({\bf q},t) = \frac{1}{N} \sum_{j=1}^N \sum_{k=1}^N \left\langle 
\exp[i {\bf q} \cdot ({\bf r}_j(t)-{\bf r}_k(t))] \right\rangle .
\label{sqt}
\ee 

\noindent
As for the pair correlation function, we perform a spherical average to
obtain $S(q,t) = S(|{\bf q}|,t)$.  In Fig.~\ref{sq}, we show the results
for the structure factor for the same set of parameters as for the pair
correlation function presented in Fig.~\ref{gr}. An advantage of $S(q,t)$
over the pair correlation function is that the local structure of the
dense phase at short-distance and the inhomogeneous bicontinuous structure
present at large scales show up in $S(q,t)$ at very different wavevectors
and thus can easily be studied independently: While the local structure
appears as a sharp peak near $q \approx 2 \pi / \sigma \approx 6$, the
large domains at larger scale produce a signal at much lower wavevectors,
and it is this low-$q$ range which is shown in Fig.~\ref{sq}.

As reported in previous work on spinodal decomposition in
fluids~\cite{koch,yamamoto}, the static structure factor develops at
low wavevector a peak whose maximum intensity, $S^\star(t)$, grows
and whose peak position, $q^\star(t)$, moves to lower $q$ as time
increases. The peak position directly reveals the typical domain size
in the phase separating system, $L(t) \approx 2 \pi / q^\star(t)$, and
the $q$-dependence near the peak can be adjusted using the following
empirical formula~\cite{furukawa}:

\be 
S(q,t) = S^\star 
\frac{3 (q/q^\star)^2  }{ 2 + ( q/q^\star )^6,  },
\label{furu}
\ee

\noindent
where the numerical factors are chosen such that
$S(q=q^\star,t)=S^\star$. This formula interpolates in a simple manner
between the expected quadratic behavior at low $q$, $S(q\ll q^\star,t)
\propto q^2$, and Porod's law describing the structure of the
interfaces at larger $q$, namely $S(q \gg q^\star,t) \propto q^{-4}$,
in a three dimensional space. We have used Eq.~(\ref{furu}) to fit the
data shown in Fig.~\ref{sq}, which gives us direct access to the growing
length scale $L(t)$ characterizing the spinodal decomposition.

We have found that this approach is much more reliable to obtain a
quantitative estimate of the average domain size than using the function
$g(r,t)$, since the large-scale signal is better resolved in Fourier
than in real space. Also, we have noticed that fluctuations are less
pronounced in $S(q,t)$ than in $g(r,t)$, which implies a reduced numerical
effort. The drawback of this simple measurement of the domain size is
however readily observed in Fig.~\ref{sq}. Since with increasing time the
peak position shifts to lower $q$, for large $t$ the peak appears at the
border of the accessible wavevector range, which is bounded at low $q$
by the system size, i.e.  $q \geq \frac{2 \pi}{L}$. This is paradoxical
at first sight, because the snapshots shown in Fig.~\ref{movie} seem to
indicate that even at large times the average domain size remains quite
a bit smaller than the box size.

The reason for this is that the behavior of the structure factor
is, just as for the pair correlation function, strongly dominated by
the largest domains in the system.  Therefore, despite the several
advantages mentioned above for the structure factor, it suffers from
the practical drawback that accurate measurements of this two-point
correlation function require system sizes that are considerably larger
than the {\it typical} domain size. To circumvent this difficulty, we 
have turned to a slightly more complicated observable, as we now describe.

\subsection{The chord length distribution}

The definition of the coarse-grained density field, given by
Eq.~(\ref{barrho}), allows to locate the position of the interfaces
separating the two phases in the course of the phase separation process.
This information can then be used to measure the distribution of the
domain size in the bicontinuous structures.

\begin{figure}
\includegraphics[width=85mm]{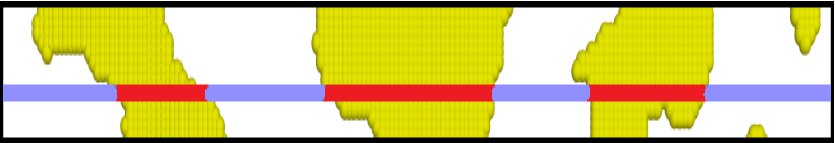}
\caption{\label{chord2} 
Chords are defined by the intersection of straight lines with the
interfaces in the phase separating system. In this bidimensional example
the red segments belong to the dense phase (yellow areas) and their
length give the chord length of the dense phase.}
\end{figure}

To this end, we measure the so-called chord length distribution (see
Fig.~\ref{chord2})~\cite{levitz,gubbins}.  We define chords by two
consecutive intersections of a straight line with the interfaces present
in the system.  In practice, we measure chords along the three axis of the
lattice used to coarse-grain the density, and measure the length $\ell$ of
the segments belonging either to the gas or to the dense phase.  For this
one has of course to take into account the periodic boundary conditions.

\begin{figure}
\includegraphics[width=85mm,clip=true, trim= 1mm 15mm 1mm 15mm]{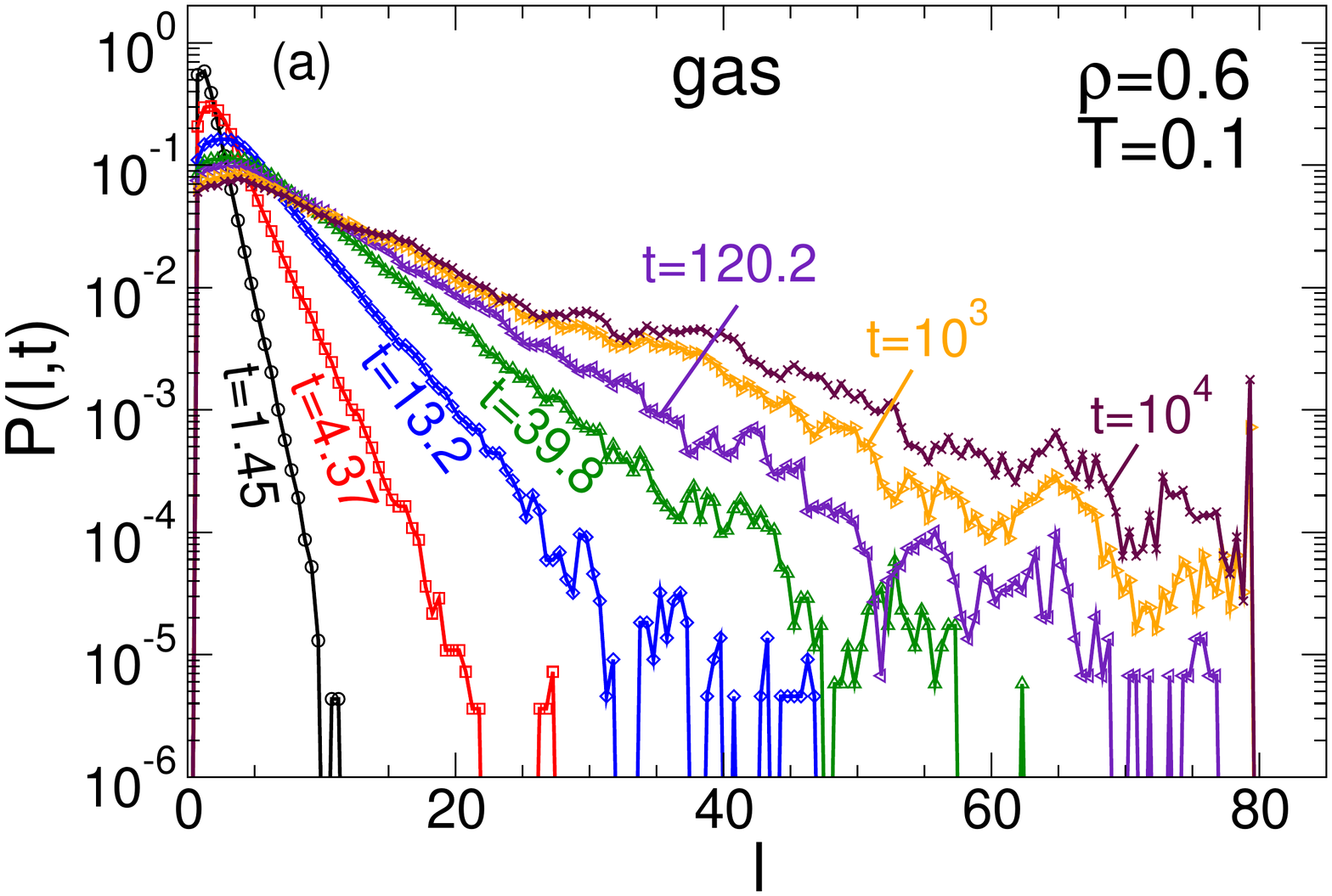}
\includegraphics[width=85mm, clip=true, trim= 1mm 20mm 1mm 25mm]{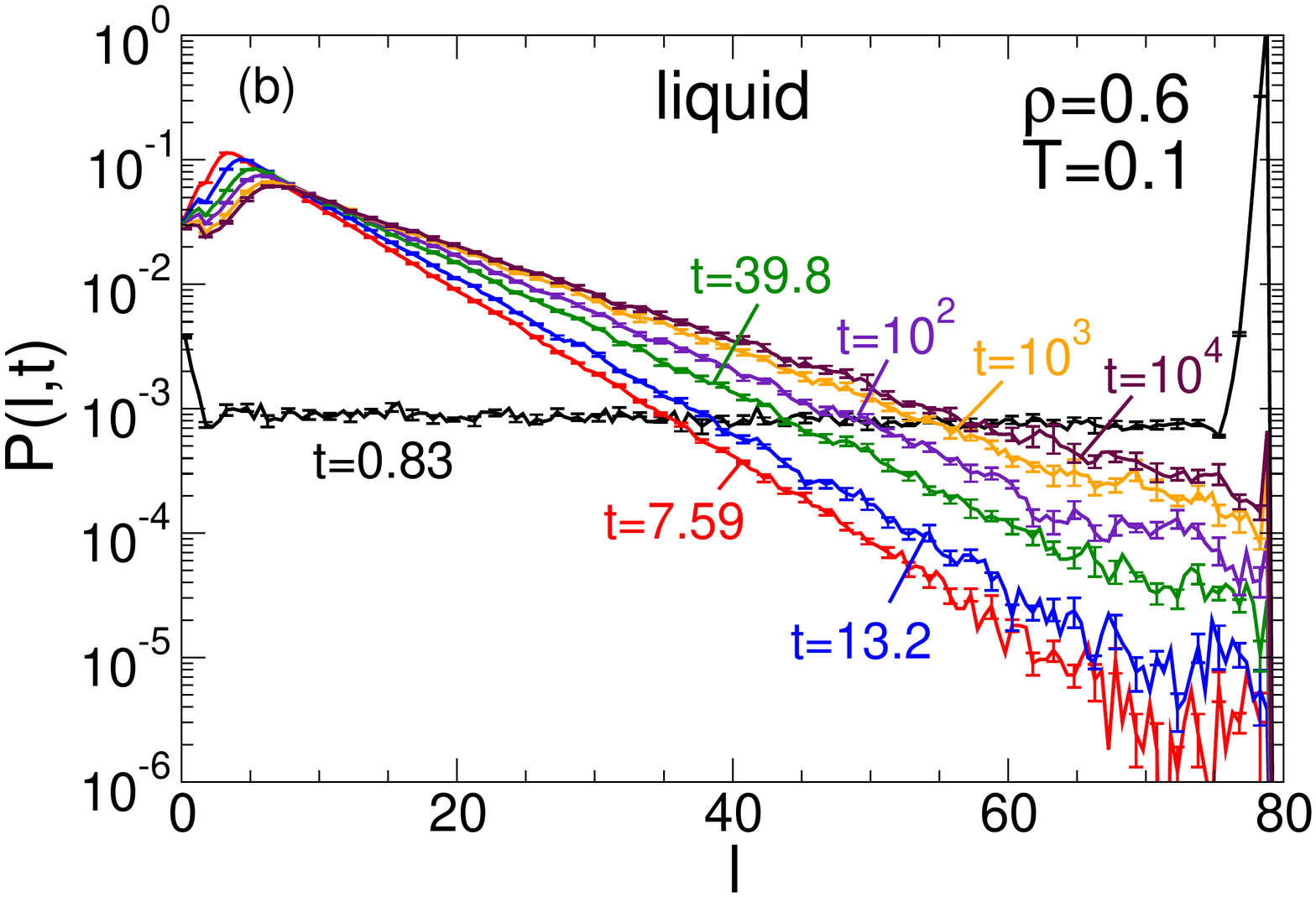}
\vspace*{-5mm}
\caption{\label{chord} 
Time evolution of the chord length distributions 
measured in the gas phase (a) and in the dense phase (b)
after a quench to $\rho=0.6$ and $T=0.1$. } 
\end{figure}

By repeating this measurement over the entire lattice used to determine
the coarse-grained density, we obtain the distribution of chord lengths
$P(\ell)$, either for chords in the gas phase, or for chords in the
dense phase.  Representative results for the time evolution of these
distributions after a quench to $\rho =0.6$ and $T=0.1$ are shown in
Fig.~\ref{chord}. These data indicate that, apart the extremely short
times when the bicontinuous structure with well-defined interfaces
has not yet developed, the two distributions are remarkably similar.
They show a maximum, which corresponds to the most probable chord length
in each phase, and beyond this length they decay asymptotically with an
exponential tail, as observed in many porous media~\cite{levitz,gubbins}.
The sharp peak located near $\ell = L/2$ for the latest time in
Fig.~\ref{chord} is due to finite size effects. Such a clear signature
is useful, since it allows to distinguish between measurements that are
affected by finite size effects from those which are not.

While either the location of the maximum or the inverse of the slope of
the exponential decay can be used as a good quantitative definition of the
average domain size, we have decided to gather all the information
stemming from the chord length distribution into one single number, and
defined the average domain size as the first moment of the distribution:

\be
L(t) = \int_0^\infty d \ell \, P(\ell,t) \ell. 
\label{domainsize}
\ee

In addition, although equivalent at long times, we find that the
distribution of chord length in the gas phase yields more accurate results
at short times than the one of the dense phase, see Fig.~\ref{chord}.
Therefore, in the rest of this paper we shall use Eq.~(\ref{domainsize})
for the chord length distribution in the gas phase as our quantitative
determination of the average domain size characterizing our phase
separating structures.

\section{Temporal evolution}

\label{temporal}

In this subsection, we analyze the temporal evolution of 
the phase separation process, and study how the kinetics depends
on the state point chosen for the quench, focusing in particular 
on the influence of temperature.

\subsection{Finite size effects}

One of the central question we wish to answer is whether or not the phase
separation kinetics is arrested at sufficiently low temperatures.  In the
previous sections we have shown that decreasing $T$ does indeed lead
to a strong slowing down of the relaxation dynamics. Before one starts
to characterize this slowing down in a more quantitative manner it is,
however, important to recall that this relaxation dynamics does depend
to some extent on the size of the system and hence one has to
check the influence of these finite size effects. In particular, we find
that coarsening stops earlier if the systems size is small. The reason
for this is that the interfaces are frustrated by the periodic boundary
conditions which constrain and slow down their motion.  This remark
is also experimentally relevant for studies of phase separation in
confined geometries~\cite{confined}.  Therefore, before concluding on the
possibility of kinetically arrested phase separations, it is important
to make sure that our results do not depend crucially on the chosen
system size.

\begin{figure} 
\includegraphics[width=90mm]{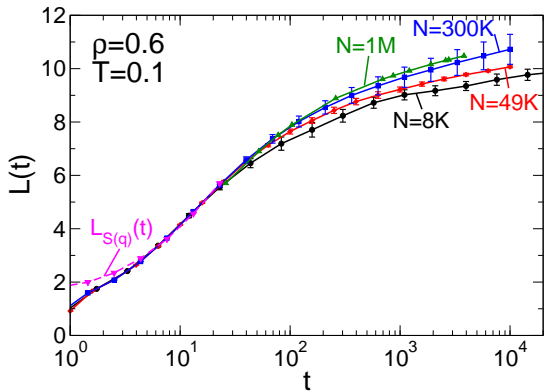}
\vspace*{-10mm}
\caption{\label{fss} Influence of system size on the growth of the
average domain size for a quench to $\rho=0.6$ and $T=0.1$, and system
size between $N=8 \cdot 10^3$ and $N=10^6$. The data for $L_{S(q)}(t)
= 2\pi / q^\star$ are deduced from the analysis of the structure factor
shown in Fig.~\ref{sq}. Note that this data has been multiplied by 0.38
in order to match the length scale $L(t)$ at intermediate times.}
\end{figure}

To this end, we have performed a systematic study of the influence
of a finite system size on the growth of the average domain size.
Some of these results are presented in Fig.~\ref{fss} for a quench
to $\rho=0.6$ and $T=0.1$.  These results confirm that the average
domain length reached at a given time after the quench increases with
increasing the system size.  However, we find that this effect does
not influence the results in a strong manner.  For the particular case
shown in Fig.~\ref{fss}, the domain size increases by about 10~\% when
$N$ increases by more than 2 orders of magnitude. Furthermore we find
no $N$ dependence within the error bars for the final sizes shown in
Fig.~\ref{fss}. Therefore we have decided to perform most of our studies
using $N = 3 \cdot 10^5$, as a compromise between a very large system,
and a broad time window in which the dynamics can be probed.

Another finding documented in Fig.~\ref{fss} is that the growth
of the length scale extracted from the chord length distribution,
Eq.~(\ref{domainsize}), matches the one obtained from the length scale
extracted from the dominant wavevector $q^\star$ in the static structure
factor, Eq.~(\ref{furu}). In the graph we have multiplied the latter
by a constant factor 0.38 and find that at short and intermediate times
the two curves do indeed track each other.  Thus one can conclude that
the chord length distribution represents an efficient and accurate way
of extracting the average domain size in phase separating systems.

\subsection{Growth of domain size}

\begin{figure}
\includegraphics[width=85mm, clip=true, trim= 1mm 15mm 1mm 10mm]{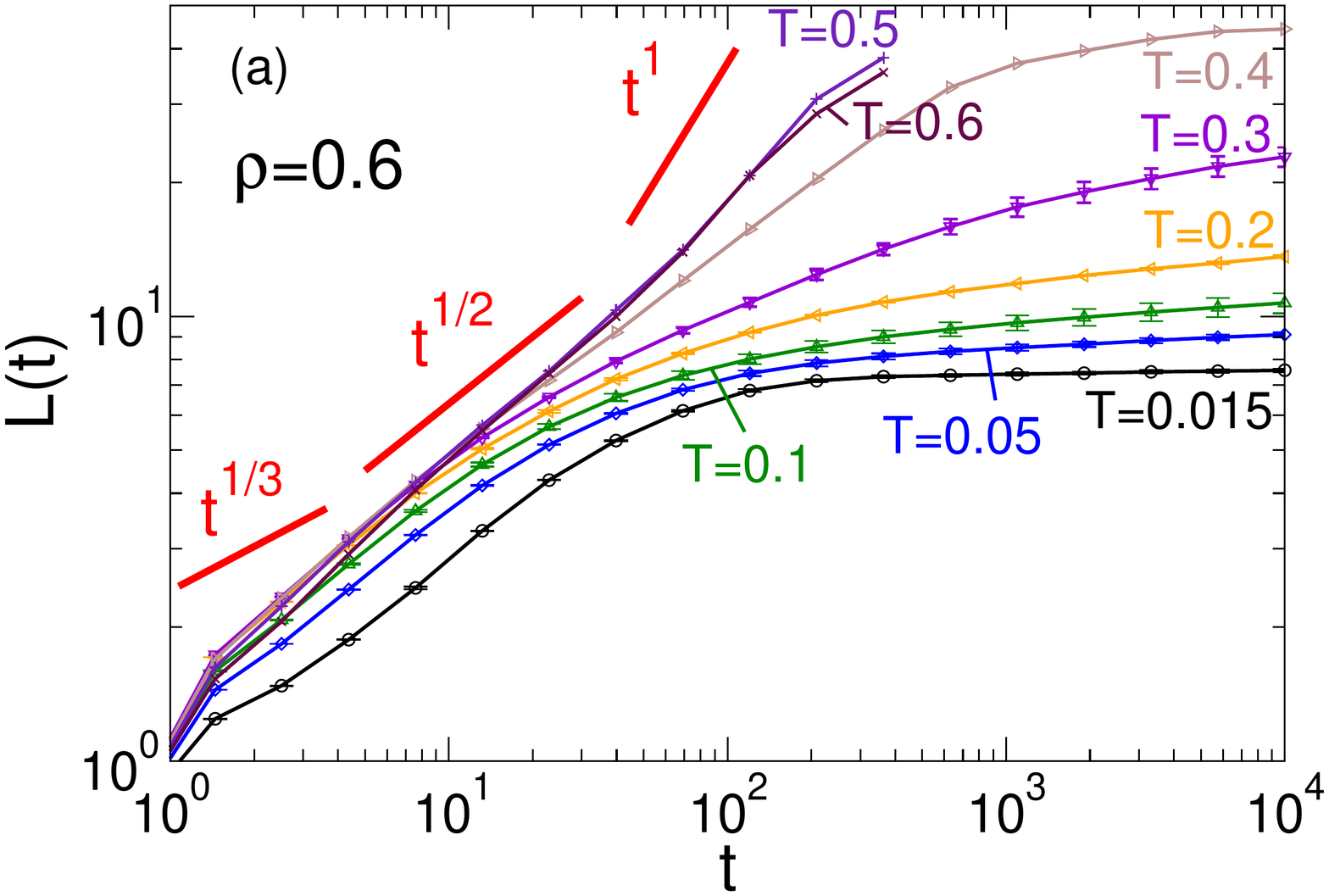}
\includegraphics[width=85mm,clip=true, trim= 1mm 15mm 1mm 25mm]{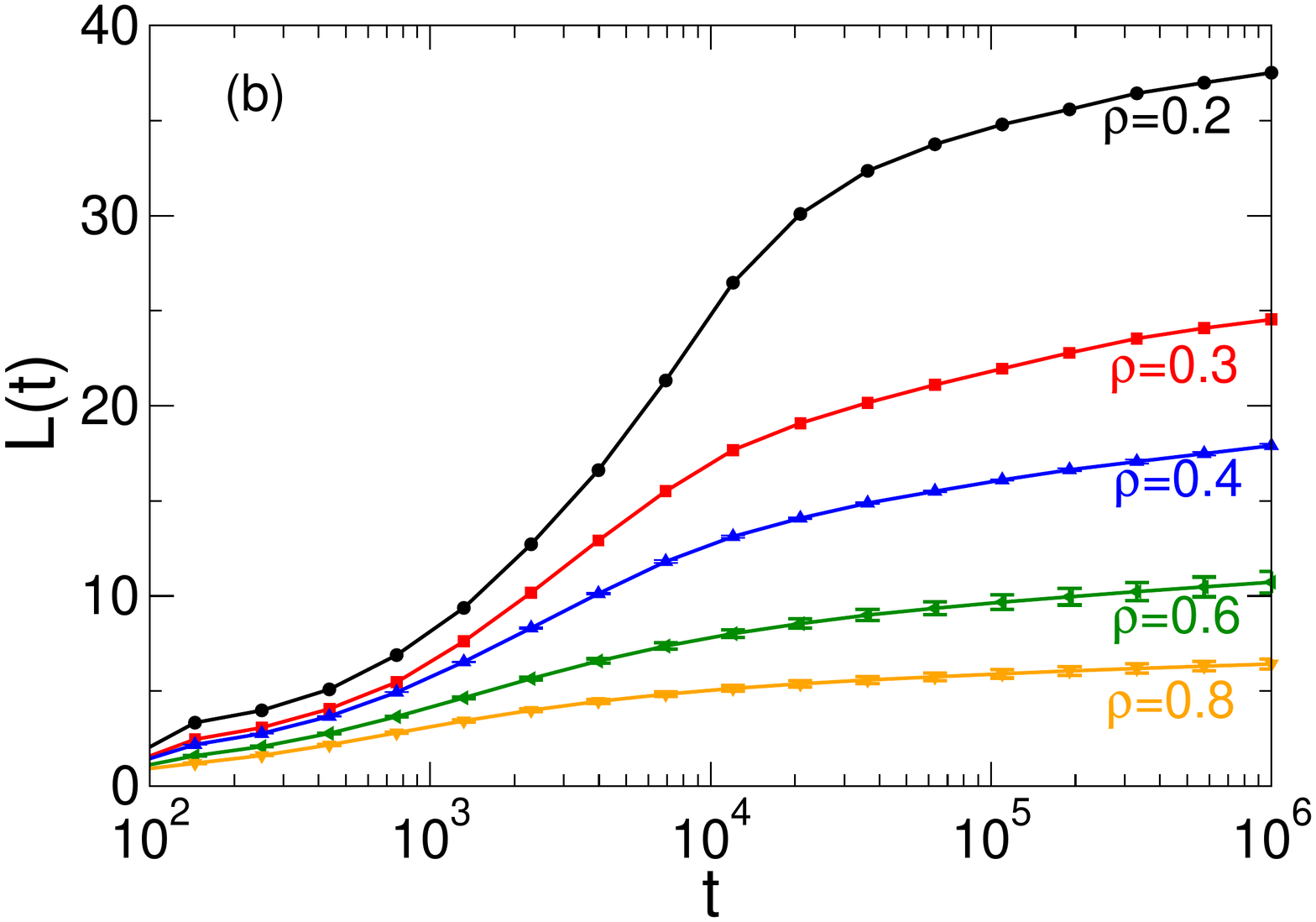}
\vspace*{-7mm}
\caption{\label{growth} a) Influence of temperature on the domain growth
for $\rho=0.6$. Various power-laws are included as well. b) Influence
of the density on the domain growth for $T=0.1$.}
\end{figure}

We now study how the density and temperature for the quench influence
the kinetics of the phase separation process. Our numerical results are
summarized in Fig.~\ref{growth}.

We first discuss the influence of temperature for a given density,
$\rho=0.6$, see Fig.~\ref{growth}a.  For a relatively shallow
quench, $T=0.5$, the data shows an upward curvature in
this double logarithmic representation before it saturates at
long times at a system size dependent value, indicating complete
phase separation. For relatively early times, the domain growth
is approximated well by $L(t) \sim t^{1/2}$, crossing over to a
faster growth at larger times before the saturation.  Such an apparent
square root time dependence has been found in earlier work on the
liquid-gas phase separation~\cite{koch,yamamoto,velasco}.  Its physical
interpretation is that it represents an effective power-law growth~\cite{lebo}
interpolating between two regimes, $L(t) \sim t^{1/3}$ at short
times followed by $L(t) \sim t$ at longer times, which are theoretically
expected power-laws controlling spinodal decomposition at early and
late times~\cite{furukawa2}.  These regimes have indeed been separately
observed in specifically dedicated simulations~\cite{cates}.  Whereas the
first regime corresponds to a surface-tension driven domain growth limited
by particle diffusion, the second one is observed when hydrodynamics
becomes relevant.  We note that our data in Fig.~\ref{growth} are
consistent with this scenario, but do not exhibit convincing power-law
regimes over broad and distinct time windows.

The situation for deeper quenches at lower temperature is more
unusual. For $T \leq 0.3$ we observe at intermediate times again an
algebraic domain growth with an exponent around 1/2. Since, however,
this $t-$dependence crosses over to one that is significantly slower, one
cannot argue that the exponent 0.5 is related to a cross-over behavior
to the hydrodynamic regime. In fact, due to the very large viscosity
of the liquid at low $T$ it must be expected that hydrodynamics ceases
to play a role~\cite{tanaka}. What is a somewhat surprising is that
at long times the growth is even slower than the usual $t^{1/3}$ law.
This indicates that at low temperatures surface tension is no longer
the main mechanism that drives the coarsening process when the domain size
becomes large and that instead a different coarsening regime sets in. This
result is in agreement with the snapshots presented in Fig.~\ref{gels}
that show that at low temperatures the interface can be rather rough
(see also Fig.~\ref{meca}).

When temperature becomes very small, $T \leq 0.2$, the data in
Fig.~\ref{growth} indicates that the domain growth at long times is not
well described by a power-law dependence, as the curves appear to be
bent in this log-log representation. This indicates that at these low
temperatures the growth is logarithmic, a functional form that is found
quite often in glassy systems.  For very low temperatures, $T=0.015$
(which is about 1~\% of the critical temperature), the domain size
ceases to grow at long times and becomes nearly constant within our
statistical accuracy.

The observation of very slow domain growth at low temperature is quite
generic, as demonstrated in Fig.~\ref{growth}b, where the density is
varied for a constant low temperature $T=0.1$. The data for $\rho =0.2$
up to $\rho=0.8$ basically follow the same time dependence, the rapid
initial domain growth becoming logarithmically slow at long times. In
contrast to temperature, density has a relatively simple influence on
the typical domain size at long times, since to a first approximation
denser systems have just smaller domains (see Fig.~\ref{movie}).

The main conclusion of this section is that for deep quenches, i.e. below
the glass temperature of the dense phase, the nature of the coarsening
process at long times becomes qualitatively different from the standard
liquid-gas phase separation. Instead of the usual power-laws we observe
a logarithmically slow domain growth, and this domain growth is only
fully arrested in the limit of vanishing temperature. This suggests that
thermal fluctuations remain relevant and control the slow domain growth
during the glass-gas phase separation.

This qualitative change has two important consequences: First, it
indicates that the microscopic mechanisms governing the phase separation
are different at low temperature from the standard surface tension
driven diffusive dynamics observed for shallow quenches.  This will be
the subject of Sec.~\ref{dynamics} below.  Second, it suggests that the
reported observation of gels formed by ``kinetically arrested'' spinodal
decomposition in colloidal systems~\cite{lu,cardinaux} is likely only
an approximation (albeit a physically relevant one) resulting from the
combination of deep quenches and short observation timescales.

\subsection{Energy density and area of interfaces}

Before discussing the details of the microscopic dynamics at low
temperature, we present two additional observables that are useful to
characterize the dynamics of the system: The energy density and area
of interfaces.

\begin{figure}
\includegraphics[width=85mm, clip=true, trim= 1mm 20mm 1mm 10mm]{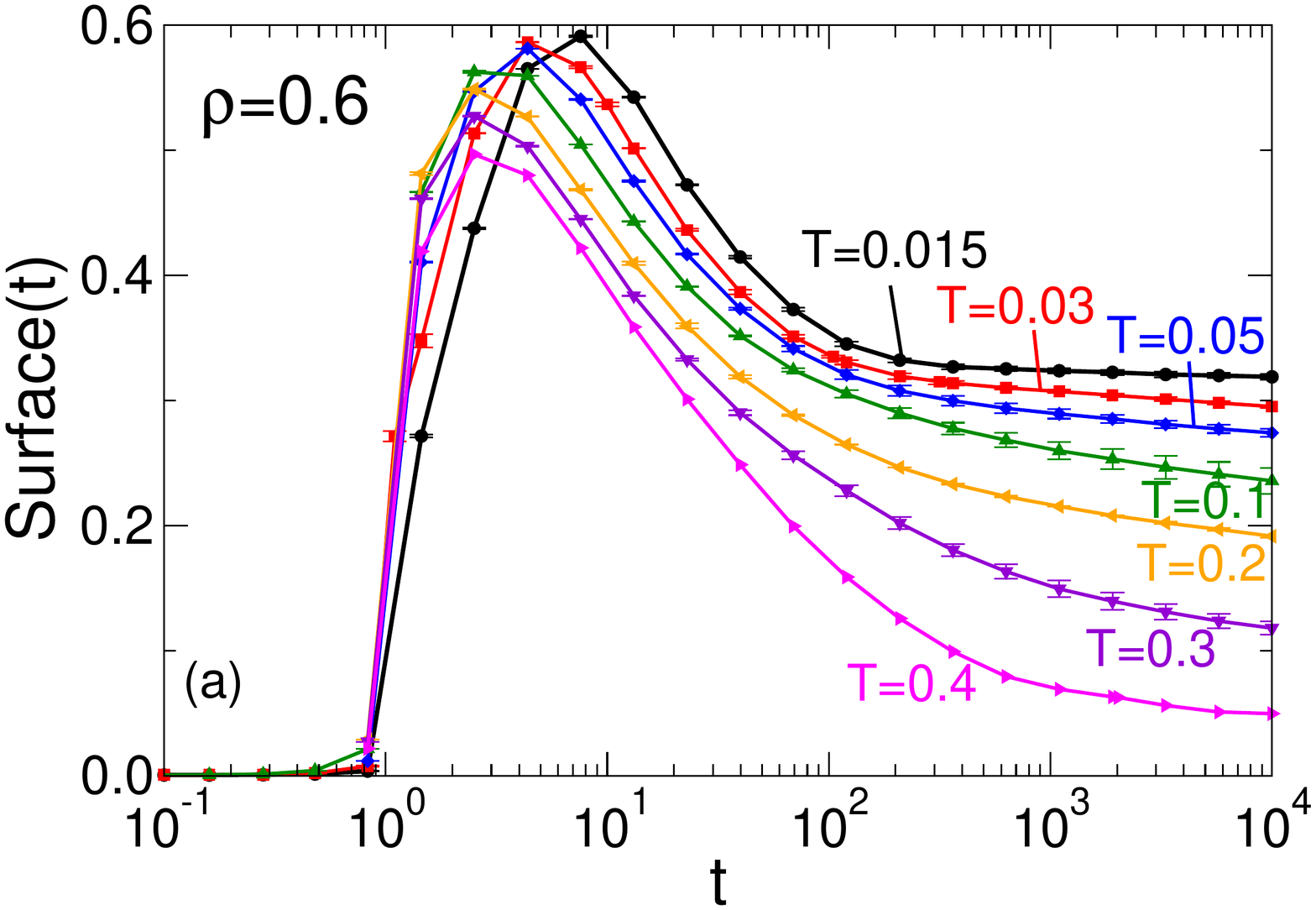}
\includegraphics[width=85mm, clip=true, trim= 1mm 15mm 1mm 20mm]{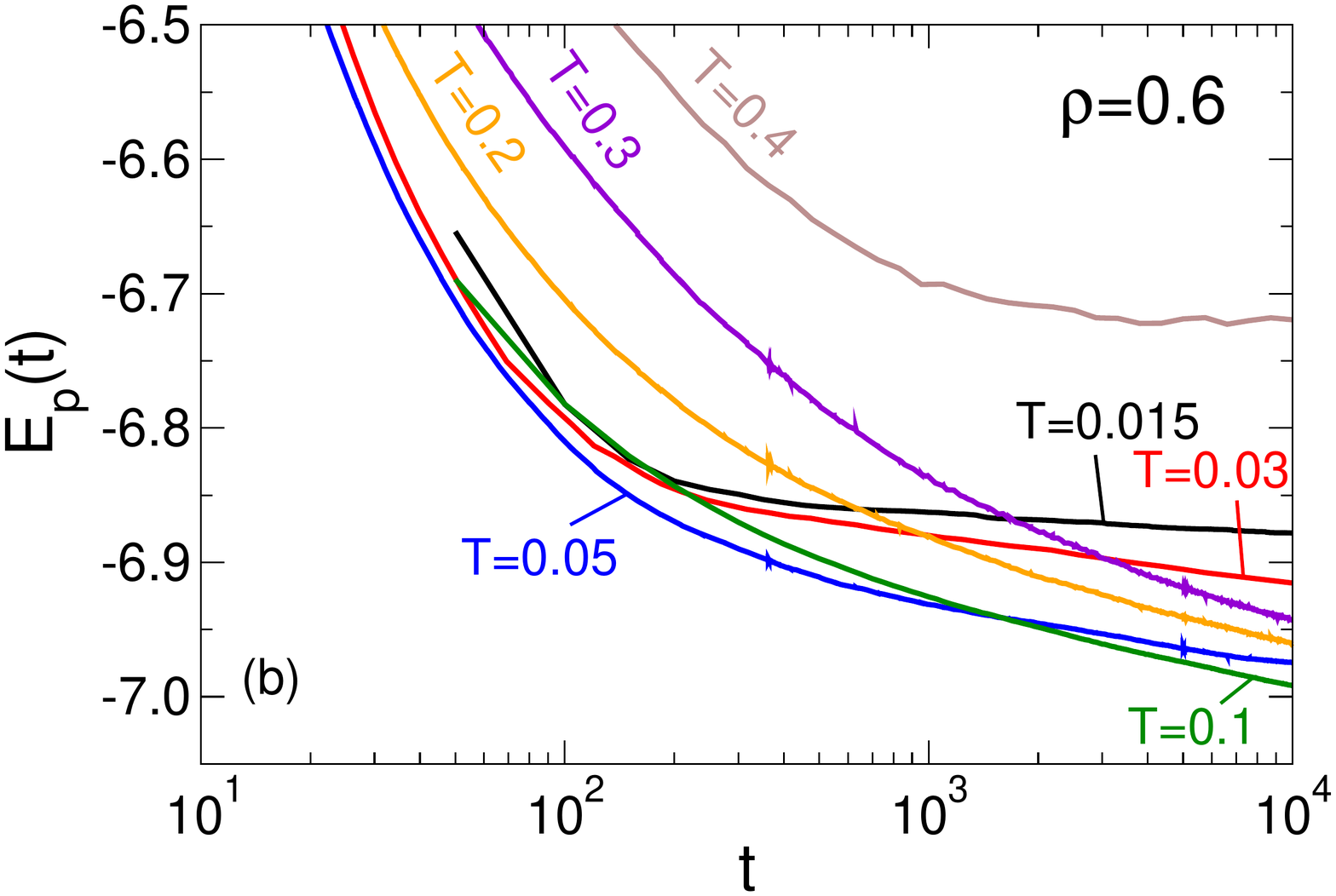}
\vspace*{-6mm}
\caption{\label{growth2} Time evolution of the surface of the interface
(a) and potential energy (b) during the phase separation at $\rho=0.6$
and various temperatures.}
\end{figure}

The time evolution of these two quantities is shown in Fig.~\ref{growth2}
for the density $\rho=0.6$ and various temperatures. The area of
the interface is here expressed as the fraction of the lattice points that are
considered as interface, see the coarse-graining procedure described
in Sec.~\ref{subcoarse}. Both observables are indicators related to the
amount of interfaces in the system, and therefore on how far the phase
separation process has advanced. Indeed, the energy density or the amount
of topological defects are sometimes used as quantitative indicators of
the average domain size during coarsening processes~\cite{bray}.

At short times the $t-$dependence of these two quantities is of
course very different: The energy density decreases rapidly since it
is dominated by the bulk behavior of the fluid which must thermalize
at low temperature after the sudden quench from the high temperature.
By contrast, because the system must first create a large quantity of
interfaces right after the quench from a homogeneous initial state, the
amount of interfaces is non-monotonous in that it first increases rapidly
at early times, before the coarsening process starts and the surface
decreases again, see Fig.~\ref{growth2}a.

For times $t > 10$ the evolution of these quantities is more similar, and
follows from similar physical considerations. At large times, the area of
the interface decreases slowly as a result of the coarsening process which
eliminates small domains and generates larger ones. The energy density
has a more complicated behavior because it receives contributions from
both the energetically costly interfaces as well as from the bulk of
the dense domains. Since both contributions decrease slowly with time,
the energy density also decreases with increasing $t$. If one assumes
that interfaces dominate the time dependence at long times, then the
energy density should display a time dependence that is very similar
to the one of the fraction of interfaces, as confirmed by the data in
Fig.~\ref{growth2} at high $T$. However, since we have seen that at
low temperatures the size of the domains is not governed by the surface
tension (cf. discussion on Fig.~\ref{growth}) one can expect that at low
$T$ the time dependence of the energy and of the surface are not the same.

The time dependence of the surface as well as of the energy
depends strongly on temperature in that the relaxation becomes
slower, in agreement with our findings regarding the size of the
domains. Figure~\ref{growth2}a shows that the amount of surface at a
given (large) time increases monotonically with decreasing temperature,
showing that at low $T$ the system has more smaller domains and their
surface is rougher. At the lowest temperatures the time dependence of
the surface is compatible with a logarithmic decay, in agreement with
our results on the growth of the domain sizes.

The increasing fraction of small domains at low temperatures has an
implication on the value of the energy at a given (large) time: Since
the energy of the dense phase decreases with $T$, one finds that for
intermediate temperatures the energy at large $t$ decreases if $T$
is lowered. However, at even lower $T$, the system starts to have
so many small domains that are rough and that cost energy, that the
overall energy starts to increase again, leading to a non-monotonic
$T-$dependence of that quantity (see Fig.~\ref{growth2}b). 

\section{Intermittent dynamics at low temperatures}

\label{dynamics}

In this section we provide evidence that the qualitative change in the
growth law at low temperatures is also accompanied by a qualitative
evolution of the microscopic mechanisms driving the coarsening dynamics.

\subsection{How domains coarsen}

If temperature is not very low, the microscopic dynamics of coarsening
is well understood. At early times of the spinodal decomposition,
a bicontinuous structure emerges rapidly, which is characterized by a
well-defined length scale that reflects the intrinsic instability of the
homogeneous system after the quench into the coexistence region. Moreover
this bicontinuous structure is characterized by curved interfaces that
store a large amount of potential energy. In this case surface tension
is the main driving force for the phase separation process that follows
the spinodal instability, and domains coarsen in order to reduce the
curvature of the interfaces and their total area. At the microscopic
scale, this process can proceed because particles can easily move within
the dense phase in response to this driving force.

At low temperature, we observe that surface tension becomes unable
to advance the coarsening process, because the dense phase is now an
(aging) glassy material that has a very high viscosity and is in fact
visco-elastic~\cite{kob_97,subdiffusiondjamel}. As a consequence,
surface tension is no longer able to relax in a significant manner the
curved interfaces formed during the phase separation process.

In a recent experimental work on attractive colloids~\cite{paddy},
it has been found that particles located near the interface of the
bicontinuous structure have a mobility which is larger than the one of
particles in the bulk of the domains. Therefore it has been concluded
that surface-enhanced mobility provides an important contribution to the
dynamics for deep quenches~\cite{paddy}. We have investigated whether
this effect is also relevant for our simulations. First of all we point
out that, due to energetic considerations, it is more favorable for the
system to put the A particles at the interface (since the A-B interaction
is stronger than the A-A interaction). Thus such a micro-segregation
would {\it a priori} give rise to an enhanced mobility of the particles
at the interface. However, despite this we have not found that for deep
quenches particles at the interfaces are significantly more mobile than
the ones inside the dense phase.  This result is in fact consistent with
earlier studies of the present binary Lennard-Jones in inhomogeneous
geometries~\cite{depablo}. While particles near surfaces are in general
indeed more mobile than particles in the bulk at equivalent thermodynamic
conditions, one should also notice that this dynamical difference is usually
only relevant in the narrow range of temperatures which corresponds to the
interval between bulk and surface glass temperatures, where bulk diffusion
is already arrested while surface diffusion is not.  The quench depth
should therefore be specifically adjusted to have conditions for which
surface-enhanced diffusion is as effective as it seems to be the case
in the experiments of Ref.~\cite{paddy}.

\begin{figure}
\includegraphics[width=85mm]{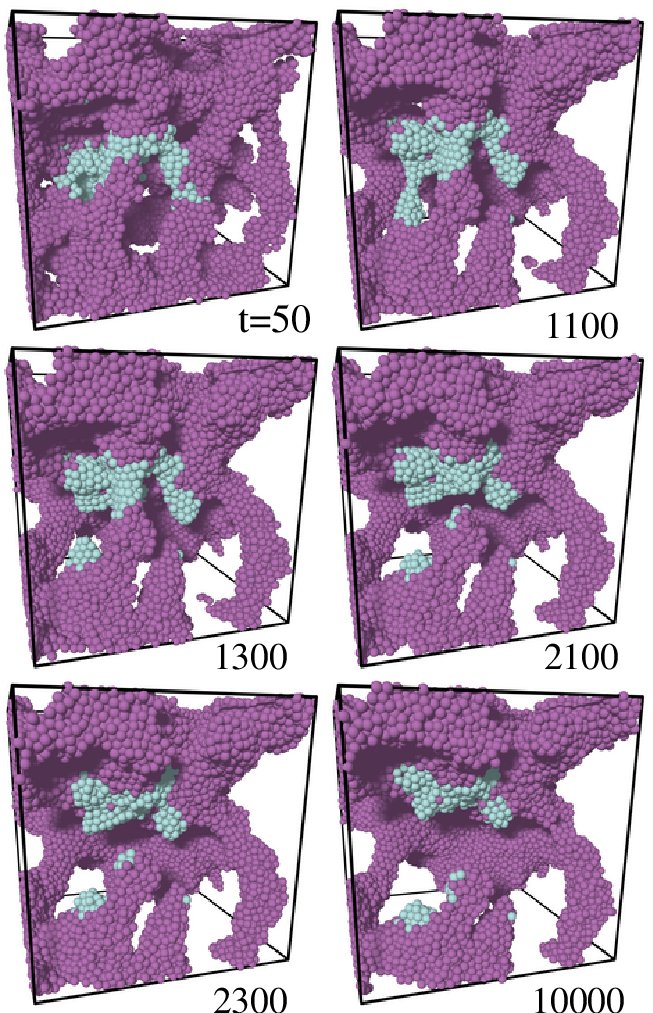}
\caption{\label{meca} Time series showing the breaking of two thin
domains in a quench at $\rho=0.4$ $T=0.1$.  Only a small fraction (16 \%)
of the entire system is shown for clarity, and about 3000 particles with
the largest mobility are highlighted, they are located near the breaking
points. The first breaking occurs between $t=50$ and 1100 and the second
one between $t=1300$ and 2100.}
\end{figure}

Although we do find in our low-temperature simulations that slow
coarsening persists, this domain growth is driven neither by surface
tension nor by interface-enhanced particle diffusion. Instead, the
complex bicontinuous structure formed at long times continues to evolve by
intermittent breaking of thin necks, which in turn allows the structure to
relax further~\cite{araki,tanakafracture}.  We illustrate this process in
the time series of Fig.~\ref{meca}, where the most mobile particles over
the considered time window are highlighted. Visual inspection shows that
for this low temperature, $T=0.1$, particles located at the interface
of a dense domain are nearly arrested. A second observation is that the
interfaces are relatively rough, which confirms that surface tension is
no longer efficient at low temperatures, and that it is unable to relax
interfaces that are very curved. A third observation is that the time
evolution of the overall structure occurs when thin domains suddenly
break, which occurs twice in this specific time series, first near $t
\approx 1100$ and then near $t \approx 2100$ in Fig.~\ref{meca}. These
sudden events are followed by a slower visco-elastic relaxation of the
structure, which eventually gives rise to an increase of the typical
domain size.

The reason for this type of relaxation behavior is related to the
fact that domains in the bicontinuous structure are under mechanical
tension. The stress field present in the glassy structure can be expected
to be very inhomogeneous because this is already a characteristic
feature in bulk amorphous solids~\cite{jlb} and the presence of a
complex geometry will certainly increase this heterogeneity. Due to
these stress inhomogeneities and the thermal fluctuations the system
will release the stored mechanical stress by breaking domains. These
events presumably occur most likely at the weak spots of the
structure, i.e.~where domains are thin or highly stressed. Once
a domain is broken, the system can relax a certain amount of
mechanical constraint, and it will reach a new metastable configuration,
until another breaking event will occur. This interpretation suggests
that it should become more and more difficult to find weak spots to
break in the system, or equivalently, that energy barriers that have
to be crossed during these events grow with time. This interpretation
naturally accounts for a logarithmic growth law for the domain size,
as is typically found in many systems with quenched disorder~\cite{book}.

Interestingly, the qualitative description of the coarsening process
occurring in our simulations at low temperatures, which results from
the intermittent release of mechanical constraints is reminiscent of the
physical scenario put forward by Cipelletti and coworkers to account for
the unusual aging dynamics observed via light scattering in a number of
soft materials~\cite{luca2,luca1,luca3}. These researchers put forward the
idea that some ``internal stress'' is stored and intermittently released
during the aging process, thus giving rise to the particular relaxation
dynamics found in these systems. This analogy suggests that it would be
very interesting to study scattering functions for the present system,
and compare the results with the behavior reported experimentally using
light scattering techniques.

\subsection{Intermittent dynamics in space and time}

\begin{figure}
\includegraphics[width=70mm]{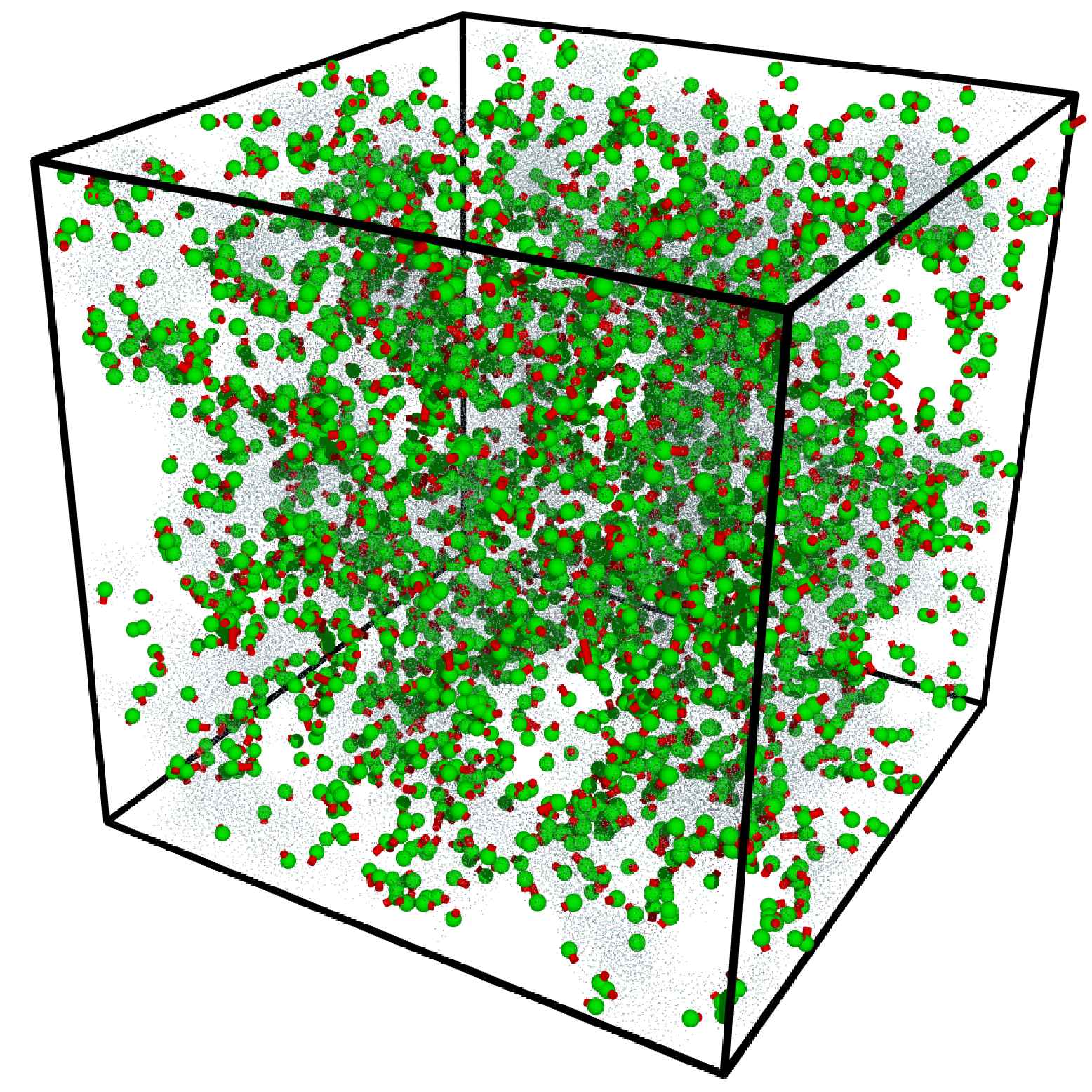}
\caption{\label{fast1} Most mobile particles (1 \% of all particles, shown
as spheres) and their displacements (shown as cylinder) for a given time
interval, $t \in [38.5, 40.5]$ after a quench to $\rho=0.4$ and $T=0.5$.}
\end{figure}

\begin{figure}
\includegraphics[width=70mm]{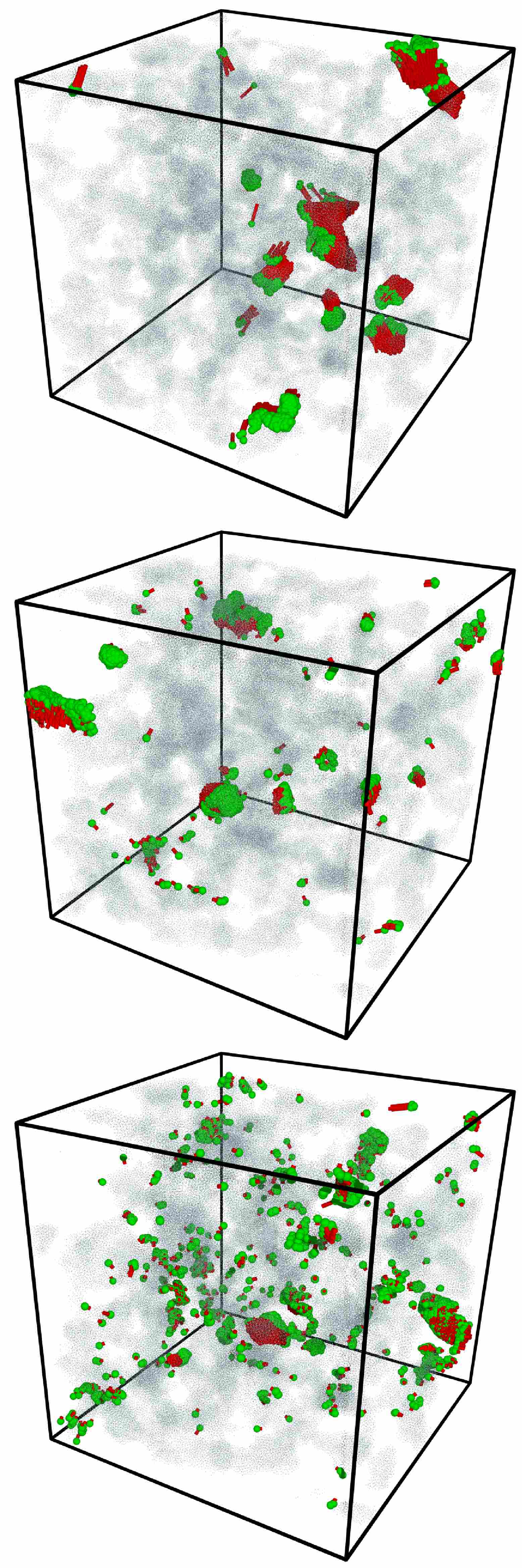}
\caption{\label{fast} 
Most mobile particles (1 \% of all particles, 
shown as spheres) and their displacements
(shown as cylinder) for three given time intervals, 
$t\in[209,363]$, 
$t\in[1905,3311]$, 
$t\in[5754,10000]$ (top to bottom), after a quench to $\rho=0.2$ and $T=0.1$.
The small dots are the remaining 99\% of the particles.}
\end{figure}

In this subsection we provide further evidence for a qualitative change
in the microscopic dynamics between shallow and deep quenches. In the
previous section we have argued that at low $T$ coarsening proceeds by
intermittent domain breaking. However, even for shallow quenches domains
grow and thin domains can break. The main difference between the
two situations is that at low temperatures, mobility is highly localized
in space and time, and domain breaking appears very rarely, whereas at
high $T$ no such localization is observed.

This difference is illustrated in Figs.~\ref{fast1} and \ref{fast},
where the 1 \% most mobile particles over different time intervals
are marked by spheres, and their displacements are represented by
a vector. The rest of the particles are represented by small dots.
The time intervals are chosen so that the images are taken for comparable
evolutions of the average domain size.  As a consequence, the times are
much shorter at the high temperature (Fig.~\ref{fast1}) than at the lower
one (Fig.~\ref{fast}), and time windows become broader with increasing
time in Fig.~\ref{fast}.

At high temperature ($T=0.5$, Fig.~\ref{fast1}) the most mobile
particles appear anywhere in the system (both at interfaces and in
the bulk domains where mobility is high) and their displacements are
essentially uncorrelated. This picture is representative of shallow
quenches and indicates that dynamic heterogeneity at the particle scale
is not very relevant for ordinary phase separations.

This behavior is in strong contrast with the one found at low temperature,
$T=0.1$ in Fig.~\ref{fast}. In this figure we show mobile particles
and their displacements over three different time intervals of the
same run, at early, intermediate, and large times.  Mobile particles
are now clearly clustered, and directly reveal the locations in the
system where the geometry of the domains has changed over a given time
interval. Furthermore one recognizes that these mobile particles also
exhibit highly correlated displacements.

Another remarkable feature is that from one image to the next, clusters
of mobile particles appear at different locations, which directly
reveals that domain breaking is a spatially and temporally intermittent
process. Finally, we notice that mobile particles almost never appear in
the interior of the domains, in contrast to the behavior found at high
temperatures. Such bulk-like relaxation appears only at very long times,
because domain breaking becomes less and less probable as the aging
proceeds.

The findings in this subsection suggest that phase separation kinetics
for deep quenches is highly intermittent in space and time, and should
therefore display a high degree of dynamic heterogeneity. It would be
interesting to apply the tools developed to study spatially heterogeneous
dynamics in glassy materials~\cite{dynhet} to quantify further the
present observations.

\section{Relation between binodal and glass lines}

\label{competition}

In this last section, we provide more details and discussion 
about the determination and behavior of the coexistence line 
at low temperatures in the phase diagram of Fig.~\ref{phase}. 

Why is this an issue? In the phase diagram of Fig.~\ref{phase},
the binodal line crosses the glass transition line near the point
$\rho \approx 1.15$ and $T \approx 0.35$. This implies that for
temperatures lower than $T=0.35$, we cannot determine the coexistence
line using standard equilibrium simulations, and we have to rely on
nonequilibrium protocols to extend the binodal down to $T=0$. Therefore
the low-temperature extension of the coexistence line changes nature at
low temperature since it is not uniquely defined anymore but is instead
dependent on the protocol.

Because of this difficulty, two experimental groups have investigated
this issue in more detail~\cite{lu,cardinaux}.  In both cases the
measurements proceed as follows: The system is first quenched into
the coexistence region where phase separation takes place, and becomes
nearly arrested on experimental timescales. Then the volume occupied by
the dense phase is determined experimentally, from which the density is
determined. Different methods have been used to measure the evolution of
this density as a function of quench depth, and two qualitatively distinct
results were found, as mentioned above in Sec.~\ref{subphase}.

For the case of our numerical simulations we have followed a similar
approach to determine the coexistence line.  After quenching  the system
to a given state point we determined the density of each phase using the
coarse-grained density field described in Sec.~\ref{subcoarse}. Following
this procedure, we obtained for each configuration both the volume occupied
by the dense phase as well as the number of particles it contains, from which
the density is easily deduced. Note that for a given temperature, the density of
the dense phase will depend in principle on both the time spent since
the quench, and on the total density of the system at which the quench
has been performed.

\begin{figure}
\includegraphics[width=93mm, clip=true, trim= 1mm 25mm 1mm 15mm]{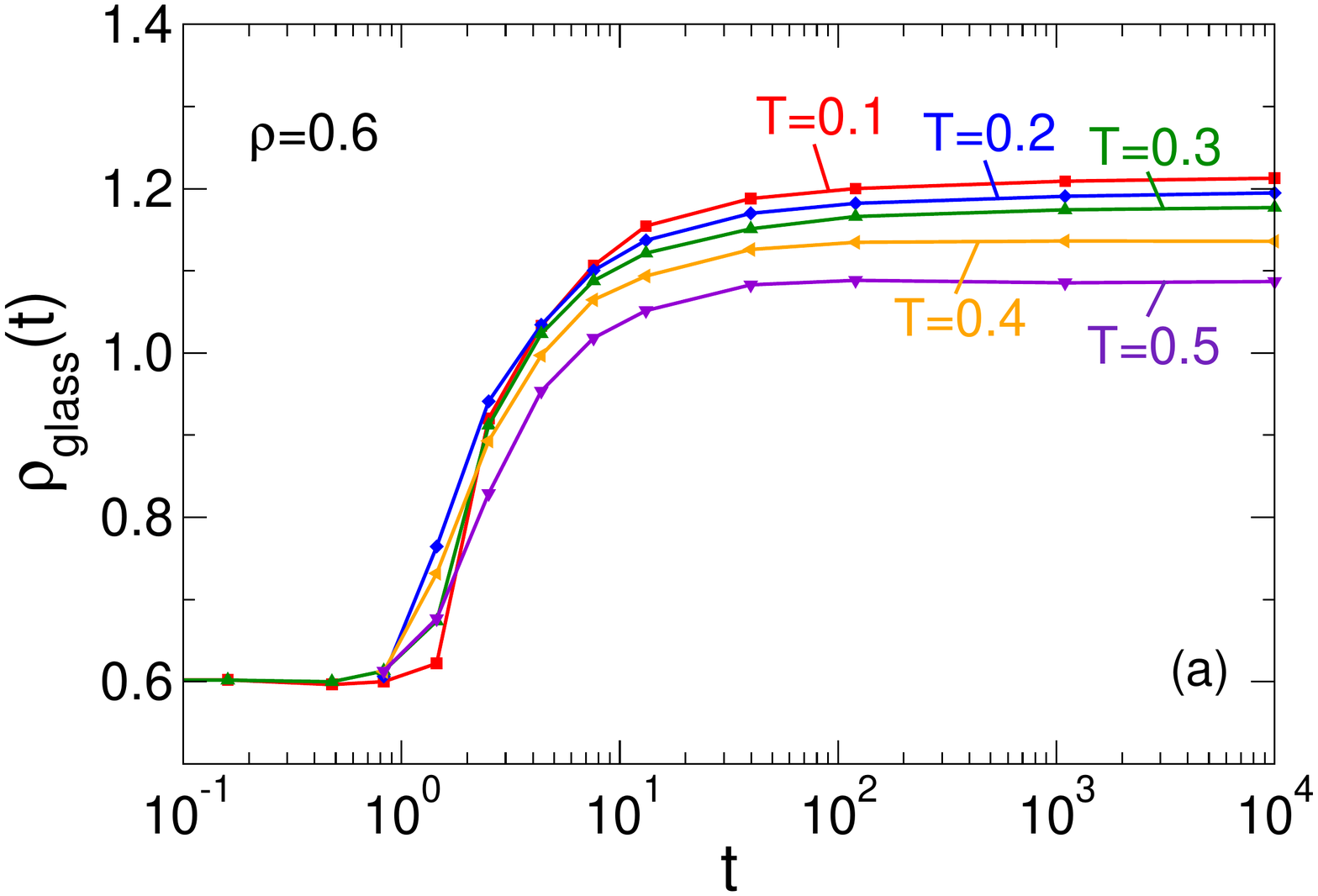}
\includegraphics[width=80mm, clip=true, trim= 1mm 15mm 1mm 20mm]{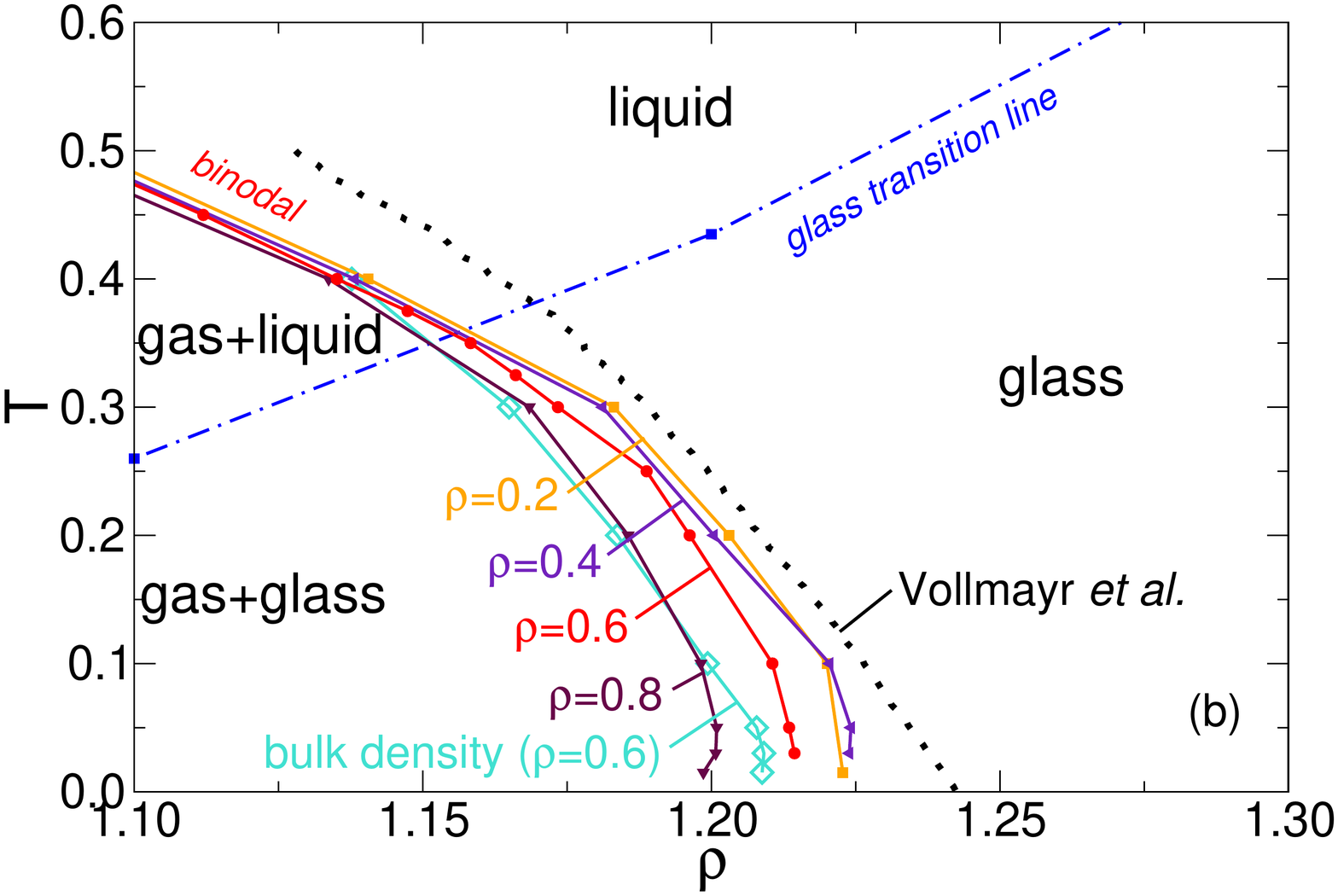}
\caption{\label{phaseZoom} 
a) Density of the dense phase during phase separation at $\rho=0.6$ and
various low temperatures.  b) Coexistence lines obtained from quenches at
various densities are reported in the phase diagram, together with the
glass transition line, and the glass density obtained in zero pressure
homogeneous glasses in Ref.~\cite{vollmayr}.}
\end{figure}

In Fig.~\ref{phaseZoom}a we demonstrate that after the quench the measured
density converges very rapidly to its asymptotic value.  This is an
important result since it shows that the heterogeneous nature of the
configurations does not preclude a quantitative determination of the
density of the glass phase. In other words, the time dependence of the
glass density is not an issue. Therefore, for a given density, here
$\rho=0.6$, we can vary the quench depth and obtain the temperature
evolution of the glass density, which we can include in the $\rho-T$
phase diagram, see Fig.~\ref{phaseZoom}b. In this representation, we
focus directly on the relevant temperature region below the intersection
with the glass transition line. We observe that the coexistence line
determined using quenches at $\rho=0.6$ changes slightly its curvature at
low temperatures, perhaps as a result of crossing the glass transition
line, but it is clearly very different from the glass line itself.
Therefore our results are closer to the ones of Ref.~\cite{lu}, which
determine the density of the glass phase using confocal microscopy and
a postprocessing which is not very different from ours. We note that the
different results reported in Ref.~\cite{cardinaux} use a more indirect
technique to measure the glass density and the coexistence line.

In Fig.~\ref{phaseZoom}b we also document the influence of the quench
density on the measured coexistence line, varying the quench density
over a broad range between $\rho=0.2$ and $\rho=0.8$. We find that all
densities produce very similar coexistence lines, with differences in
density of about 2 \% between the two extremes, the larger quench density
producing a smaller glass density.

Finally we compare the results for the coexistence line with a very
different numerical approach.  In Ref.~\cite{vollmayr}, the present
Lennard-Jones binary mixture was used to study the influence of
cooling rates on the structure of the glass. These quenches were done
at constant pressure, which was zero. Thus, the produced homogeneous
glass configurations were adjusting their densities to maintain a zero
pressure, and these densities were recorded numerically. We have included
the temperature evolution of these densities in Fig.~\ref{phaseZoom}b as
well and we see that they compare very well with our determination of the
coexistence region.  This is expected because the dense phase in our phase
separating systems coexists with a gas phase with vanishing pressure.
This comparison seems to confirm our finding that the coexistence line
does not exhibit a reentrant behavior as a result of the crossing of
the glass transition line.

\section{Summary and conclusion}

\label{conclusion}

We have used large-scale molecular dynamics simulations to study the
influence of a temperature quench on the liquid-gas phase separation
kinetics in a Lennard-Jones fluid, and therefore the competition between
the phase separation kinetics and the glass transition occurring at low
temperature in bulk liquids. This represents therefore an example of a
viscoelastic phase separation.

Our main finding is the observation that the phase separation kinetics
changes qualitatively with decreasing temperature: The microscopic
dynamics evolves from a diffusive motion driven by surface tension for
shallow quenches, to a qualitatively different coarsening regime in which
the dynamics becomes strongly intermittent, spatially heterogeneous and
thermally activated at low temperature, leading to logarithmically slow
growth of the typical domain size.

The microscopic description of the coarsening process occurring in our
simulations at low temperatures, which results from the intermittent
release of mechanical constraints, is strongly reminiscent of the physical
scenario put forward to explain experimental and simulation results
obtained in a broad variety of soft glassy materials for which unusual
aging dynamics has been reported~\cite{luca2,luca1,luca3,suarez_2009}. In
future work, it would be interesting to compare time correlation functions
measured in numerical simulations such as ours to the outcome of light
scattering experiments performed in soft glassy materials in their
aging regime.  Such studies would allow to obtain a better understanding
how these soft glass materials are related to the gel-like structures
investigated here.

\acknowledgments

We thank D. Reichman for initially stimulating this work, and B. Coasne,
T. Gibaud, and P. Royall for useful discussions.  The research leading
to these results has received funding from the European Research Council
under the European Union's Seventh Framework Programme (FP7/2007-2013)
/ ERC Grant agreement No 306845.  W. Kob acknowledges support from the
Institut Universitaire de France.

\end{document}